\documentclass[twocolumn]{aastex631}

\usepackage{amsmath}	

\shorttitle{NIRSpec Observations of Ly$\alpha$ at $z>6.5$}
\shortauthors{Tang et al.}

\begin{document}

\title{\textit{JWST}/NIRSpec Observations of Ly$\alpha$ Emission in Star Forming Galaxies at $6.5\lesssim z\lesssim13$}

\author[0000-0001-5940-338X]{Mengtao Tang}
\affiliation{Steward Observatory, University of Arizona, 933 N Cherry Ave, Tucson, AZ 85721, USA}
\email{tangmtasua@arizona.edu}

\author{Daniel P. Stark}
\affiliation{Steward Observatory, University of Arizona, 933 N Cherry Ave, Tucson, AZ 85721, USA}

\author{Michael W. Topping}
\affiliation{Steward Observatory, University of Arizona, 933 N Cherry Ave, Tucson, AZ 85721, USA}

\author{Charlotte Mason}
\affiliation{Cosmic Dawn Center (DAWN), Niels Bohr Institute, University of Copenhagen, Jagtvej 128, 2200 K{\o{}}benhavn N, Denmark}

\author{Richard S. Ellis}
\affiliation{Department of Physics and Astronomy, University College London, Gower Street, London WC1E 6BT, UK}



\begin{abstract}
We present an analysis of \textit{JWST} Ly$\alpha$ spectroscopy of $z\gtrsim6.5$ galaxies, using observations in the public archive covering galaxies in four independent fields (GOODS-N, GOODS-S, Abell 2744, EGS). We measure Ly$\alpha$ emission line properties for a sample of $210$ $z\simeq6.5-13$ galaxies, with redshifts confirmed independently of Ly$\alpha$ in all cases. We present $3$ new detections of Ly$\alpha$ emission in \textit{JWST} spectra, including a large equivalent width (EW $=143$~\AA) Ly$\alpha$ emitter with strong C~{\small IV} emission (EW $=21$~\AA) at $z=7.1$ in GOODS-N. We measure the redshift-dependent Ly$\alpha$ EW distribution across our sample. We find that strong Ly$\alpha$ emission (EW $>25$~\AA) becomes increasingly rare at earlier epochs, suggesting that the transmission of Ly$\alpha$ photons decreases by $4\times$ between $z\simeq5$ and $z\simeq9$. We describe potential implications for the IGM neutral fraction. There is significant field to field variance in the Ly$\alpha$ emitter fraction. In contrast to the three other fields, the EGS shows no evidence for reduced transmission of Ly$\alpha$ photons at $z\simeq7-8$, suggesting a significantly ionized sightline may be present in the field. We use available NIRCam grism observations from the FRESCO survey to characterize overdensities on large scales around known Ly$\alpha$ emitters in the GOODS fields. The strongest overdensities appear linked with extremely strong Ly$\alpha$ detections (EW $>50$~\AA) in most cases. Future Ly$\alpha$ spectroscopy with \textit{JWST} has the potential to constrain the size of ionized regions around early galaxy overdensities, providing a new probe of the reionization process.
\end{abstract}




\section{Introduction} \label{sec:introduction}

Ly$\alpha$ emitting galaxies have long been used as probes of the reionization of intergalactic hydrogen (e.g., \citealt{Malhotra2004,Ouchi2010,Dijkstra2014,Jung2020,Jones2024}; see \citealt{Ouchi2020} for a review), complementing measurements from the cosmic microwave background (CMB; \citealt{Planck2020}) and deep quasar spectra \citep[e.g.,][]{Banados2018,Davies2018,Wang2020,Yang2020,Greig2022}. 
The discovery of the first robust samples of star forming galaxies at $z\simeq7-8$ in 2009 \citep[e.g.,][]{Bouwens2010,Bouwens2011,Bunker2010,McLure2010,McLure2011,Oesch2010,Wilkins2010} sparked a series of ground-based spectroscopic campaigns with Keck and Very Large Telescope (VLT) aimed at measuring the distribution of Ly$\alpha$ line strengths at $z\gtrsim6$ \citep[e.g.,][]{Stark2010,Fontana2010,Vanzella2011,Ono2012,Pentericci2014,Hoag2019,Mason2019a}. 
The goal of these early investigations was to determine if the intergalactic medium (IGM) was sufficiently neutral at $z\simeq7-8$ for its damping wing to attenuate the Ly$\alpha$ line. 
These surveys quickly demonstrated that the fraction of strong Ly$\alpha$ emitters (LAEs) in $z\simeq7-8$ galaxy samples is lower than that at $z\simeq5-6$ \citep[e.g.,][]{Ono2012,Treu2013,Schenker2014,Tilvi2014,Pentericci2018,Mason2019a}, indicating a significant reduction in transmission of Ly$\alpha$ photons at $z\gtrsim7$. 
If due to the damping wing of the IGM, the disappearance of Ly$\alpha$ emitters would point to significant neutral hydrogen (H~{\small I}) fractions ($x_{\rm HI}\gtrsim0.5$) at $z\simeq7$ \citep[e.g.,][]{Caruana2014,Zheng2017,Mason2018,Hoag2019,Whitler2020,Bolan2022,Nakane2024}. 
Similar indications of a partially neutral $z\simeq7$ IGM also emerged from comprehensive quasar surveys \citep[e.g.,][]{McGreer2015,Jin2023,Durovcikova2024}, suggesting a consistent picture for the latter half of reionization. 

The launch of \textit{JWST} \citep{Gardner2023} has extended the galaxy frontier to $z\simeq10-15$ \citep[e.g.,][]{ArrabalHaro2023a,ArrabalHaro2023b,Bunker2023a,Curtis-Lake2023,DEugenio2023,Fujimoto2023b,Carniani2024,Castellano2024,Chakraborty2024,Witstok2024c}, offering the potential to apply the Ly$\alpha$ test to the early stages of reionization. 
Deep \textit{JWST} spectroscopy offers several distinct advantages with respect to earlier ground-based investigations. 
The absence of sky lines and atmospheric absorption has improved redshift completeness and reliability of flux measurements. 
The improved spectroscopic sensitivity and confident rest-frame optical redshifts allow meaningful Ly$\alpha$ measurements in $z\gtrsim7$ sources with continuum magnitudes $\gtrsim10-50\times$ fainter than was possible from the ground (e.g., \citealt{Saxena2023,Chen2024}), greatly improving statistics. 
Access to systemic redshifts and Balmer line fluxes have provided estimates of the velocity profiles and escape fraction of Ly$\alpha$, offering new insight into how the IGM is modulating Ly$\alpha$ \citep[e.g.,][]{Bunker2023a,Tang2023,Tang2024a,Tang2024b,Saxena2023,Saxena2024,Chen2024,Witstok2024b}.

Early results have demonstrated the potential of \textit{JWST} to transform distant Ly$\alpha$ emitter investigations. 
\textit{JWST} spectroscopy of the $z=10.6$ galaxy GN-z11 \citep{Oesch2016} quickly revealed the capability of the Near Infrared Spectrograph (NIRSpec; \citealt{Jakobsen2022,Boker2023}) in recovering Ly$\alpha$ emission at $z\gtrsim10$ \citep{Bunker2023a,Witstok2024c}. 
Meanwhile NIRSpec observations of the Extended Groth Strip (EGS; \citealt{Davis2007}) field revealed several extremely strong Ly$\alpha$ emitters within several physical Mpc (pMpc) of previously-known bright Ly$\alpha$ emitters \citep{Tang2023,Chen2024,Nakane2024,Napolitano2024}. 
The implied Ly$\alpha$ escape fractions and equivalent widths (EWs) suggest minimal IGM attenuation, as would be expected if EGS was host to one or more large ($\gtrsim1$~pMpc) ionized line-of-sight structures at $z\simeq7-8$ \citep[e.g.,][]{Tilvi2020,Leonova2022,Tang2023,Chen2024,Whitler2024}. 
Variations in Ly$\alpha$ visibility have been found in the EGS and Great Observatories Origins Deep Survey (GOODS; \citealt{Giavalisco2004}) South fields \citep{Napolitano2024}, possibly indicating significant differences in the typical IGM ionization state at $z\simeq7-8$ in the two deep fields. 

As of summer 2024, there is now a large database of NIRSpec observations targeting $z\gtrsim6.5$ galaxies in four independent fields: Abell 2744 \citep{Bezanson2022,Treu2022}, EGS \citep{Finkelstein2024}, and GOODS-North and South \citep{Bunker2023b,DEugenio2024}. 
In this paper, we present an analysis of the Ly$\alpha$ statistics of the full public dataset, with self-consistent reduction and selections. 
We measure Ly$\alpha$ properties for $210$ galaxies at $z>6.5$, a sample that is $3-6$ times larger than previous studies using NIRSpec \citep[e.g.,][]{Jones2024,Nakane2024,Napolitano2024}. 
We seek to quantify the distribution of Ly$\alpha$ EWs and Ly$\alpha$ escape fractions at $z\gtrsim6.5$ with improved statistics relative to earlier investigations. 
Maximizing the number of independent fields is critical for these measurements given the range of IGM ionization states expected along different sightlines. 
We will use the evolving Ly$\alpha$ distributions to quantify the changing transmission of Ly$\alpha$ emission between $z\simeq5$ and $z\simeq13$, discussing potential implications for the IGM. 
This compilation will provide insight into the state of Ly$\alpha$ observations two years after the first \textit{JWST} data, giving a baseline of what remains to be done to maximize the potential of \textit{JWST} for constraining reionization with Ly$\alpha$ datasets. 
The next key step involves investigating the galaxy populations in regions that Ly$\alpha$ observations suggest are likely to be ionized. 
We will describe what early \textit{JWST} observations are revealing about the galaxy environments associated with the strongest $z\gtrsim7$ Ly$\alpha$ emitters. 

The organization of this paper is as follows. 
In Section~\ref{sec:data}, we describe the sample of galaxies at $z>6.5$ identified from publicly available \textit{JWST}/NIRSpec observations. 
We then characterize the Ly$\alpha$ EWs and Ly$\alpha$ escape fractions of galaxies in our $z>6.5$ sample and derive the statistical distributions of Ly$\alpha$ properties in Section~\ref{sec:lya_evolution}. 
Using the Ly$\alpha$ EW measurements, we infer the redshift-dependent transmission of Ly$\alpha$ over $5\lesssim z\lesssim13$ and discuss the implications for the neutral hydrogen fraction in the IGM in Section~\ref{sec:igm_transmission}. 
In Section~\ref{sec:lya_association}, we discuss the galaxy environments associated with strong Ly$\alpha$ emitting galaxies at $z\gtrsim7$. 
Finally, we summarize our conclusions in Section~\ref{sec:summary}. 
Throughout the paper we adopt a $\Lambda$-dominated, flat universe with $\Omega_{\Lambda}=0.7$, $\Omega_{\rm{M}}=0.3$, and $H_0=70$~km~s$^{-1}$~Mpc$^{-1}$. 
All magnitudes are quoted in the AB system \citep{Oke1983} and all EWs are quoted in the rest frame.


\section{Spectroscopic Sample and Analysis} \label{sec:data}

In this section, we construct and analyze the Ly$\alpha$ properties of a sample of $210$ galaxies at $z>6.5$ with publicly-available NIRSpec spectra. 
We introduce the spectra in Section~\ref{sec:spectra}, and select the $z>6.5$ sample in Section~\ref{sec:sample}. 
We describe the photometric measurements and fit the spectral energy distributions (SEDs) of our sources in Section~\ref{sec:photometry}. 
In Section~\ref{sec:new_lae}, we present three newly identified Ly$\alpha$ emitting galaxies at $z>6.5$. 
Then we present the spectroscopic measurements of the entire $z>6.5$ sample, characterizing the Ly$\alpha$ EWs, Ly$\alpha$ escape fractions, and Ly$\alpha$ velocity offsets in Section~\ref{sec:measurements}.

\subsection{\textit{JWST}/NIRSpec Spectra} \label{sec:spectra}

The NIRSpec spectra used in this work were obtained from the following public observations: 
the \textit{JWST} Advanced Deep Extragalactic Survey\footnote{\url{https://jades-survey.github.io/}} (JADES, GTO 1180, GTO 1181, PI: D. Eisenstein, GTO 1210, PI: N. L\"utzgendorf, GO 3215, PI: D. Eisenstein \& R. Maiolino; \citealt{Bunker2023b,Eisenstein2023a,Eisenstein2023b,DEugenio2024}), the GLASS-\textit{JWST} Early Release Science Program\footnote{\url{https://glass.astro.ucla.edu/ers/}} (ERS 1324, PI: T. Treu; \citealt{Treu2022}), the Cosmic Evolution Early Release Science\footnote{\url{https://ceers.github.io/}} (CEERS, ERS 1345, PI: S. Finkelstein; \citealt{Finkelstein2024}) and a Director’s Discretionary Time program (DDT 2750, PI: P. Arrabal Haro; \citealt{ArrabalHaro2023a,ArrabalHaro2023b}), as well as the Ultra-deep NIRCam and NIRSpec Observations Before the Epoch of Reionization\footnote{\url{https://jwst-uncover.github.io/}} (UNCOVER, GO 2561, PI: I. Labb\'e \& R. Bezanson; \citealt{Bezanson2022,Price2024}). 
All NIRSpec observations were performed with the multi-object spectroscopy (MOS) mode using the micro-shutter assembly (MSA; \citealt{Ferruit2022}). 
We refer readers to the above references for detailed descriptions of the NIRSpec observations. 
Below we briefly summarize these observations. 

The JADES GTO 1180, GTO 1210, and GO 3215 observations targeted the GOODS-South field, and the GTO 1181 observations targeted the GOODS-North field. 
GTO 1180 so far have observed nine different pointings, using both the low resolution ($R\sim100$; corresponding to velocity $\sim3000$~km~s$^{-1}$) prism covering a wavelength range of $0.6-5.3\ \mu$m and the medium resolution ($R\sim1000$; or $\sim300$~km~s$^{-1}$ resolution in velocity) gratings with three grating/filter pairs G140M/F070LP, G235M/F170LP, and G395M/F290LP covering $0.7-5.3\ \mu$m. 
Each 1180 pointing was observed with exposure time of $1.1-3.1$~hours for prism and $0.9-2.6$~hours for each grating. 
GTO 1181 observed nine pointings using the prism and three medium resolution gratings (G140M/F070LP, G235M/F170LP, and G395M/F290LP), and three of the nine pointings were also observed using one high resolution ($R\sim2700$; or $\sim110$~km~s$^{-1}$ in velocity) grating/filter pair G395H/F290LP. 
Each 1181 pointing took exposure time of $1.8-2.6$~hours for prism and $0.9-2.6$~hours for each grating. 
GTO 1210 observed one pointing using the prism, three medium resolution grating/filter pairs (G140M/F070LP, G235M/F170LP, and G395M/F290LP), and one high resolution grating/filter pair (G395H/F290LP), with exposure time of $27.8$~hours ($6.9$~hours) for prism (each grating). 
GO 3215 observed one pointing using the prism and two medium resolution grating/filter pairs (G140M/F070LP and G395M/F290LP), with exposure time of $46.7$, $11.7$, and $46.7$~hours for prism, G140M/F070LP, and G395M/F290LP. 

The GLASS and the UNCOVER observations targeted the field behind the lensing galaxy cluster Abell 2744 \citep{Abell1989}. 
The GLASS ERS 1324 observed one pointing using three high resolution grating/filter pairs G140H/F100LP, G235H/F170LP, and G395H/F290LP (covering $1.0-5.3\ \mu$m), with an exposure time of $4.9$~hours for each grating. 
The UNCOVER GO 2561 observed seven MSA mask configurations using the prism, with an exposure time of $2.4-4.4$~hours for each configuration. 
The CEERS ERS 1345 and DDT 2750 observations targeted the EGS field. 
ERS 1345 observed six pointings using the prism and another six pointings using three medium resolution grating/filter pairs (G140M/F100LP, G235M/F170LP, and G395M/F290LP), with four overlapped pointings observed with both the prism and gratings. 
The exposure time of each pointing is $0.9$~hour, except for one prism pointing with $1.7$~hours. 
DDT 2750 spectra observed one pointing using the prism, with an exposure time of $5.1$~hours. 
We list the details of the NIRSpec observations used in this study in Table~\ref{tab:nirspec_obs}. 


\begin{deluxetable*}{cccccccccc}
\tablecaption{Summary of NIRSpec/MSA observations used in this work. PID represents the \textit{JWST} proposal ID of each program. We list the prism and gratings used in each program and the total exposure time per pointing.}
\tablehead{
PID & Field & Pointings & PRISM & G140M & G235M & G395M & G140H & G235H & G395H \\
 & & & (h) & (h) & (h) & (h) & (h) & (h) & (h) 
}
\startdata
GTO 1180$^{\rm a}$ & GOODS-S & 9 & $1.1-3.1$ & $0.9-2.6$ & $0.9-2.6$ & $0.9-2.6$ & - & - & - \\
GTO 1181$^{\rm b}$ & GOODS-N & 9 & $1.8-2.6$ & $0.9-2.6$ & $0.9-2.6$ & $0.9-2.6$ & - & - & $2.6$ \\
GTO 1210 & GOODS-S & 1 & $27.8$ & $6.9$ & $6.9$ & $6.9$ & - & - & $6.9$ \\
GO 3215 & GOODS-S & 1 & $46.7$ & $11.7$ & - & $46.7$ & - & - & - \\
ERS 1324 & Abell 2744 & 7 & - & - & - & - & $4.9$ & $4.9$ & $4.9$ \\
ERS 1345$^{\rm c}$ & EGS & 6 & $0.9-1.7$ & $0.9$ & $0.9$ & $0.9$ & - & - & - \\
GO 2561 & Abell 2744 & 7 & $2.4-4.4$ & - & - & - & - & - & - \\
DDT 2750 & EGS & 1 & $5.1$ & - & - & - & - & - & - \\
\enddata
\tablecomments{a: Seven of the nine 1180 pointings were observed with exposure time of $1.1$~hours for prism and $0.9$~hour for each grating at each pointing, one was observed with $2.1$~hours for prism and $1.7$~hours for each grating, and the remaining one was observed with $3.1$~hours for prism and $2.6$~hours for each grating. b: Six of the nine 1181 pointings were observed with the prism and three medium resolution gratings, with exposure time of $1.8$~hours for prism and $0.9$~hour for each grating at each pointing. The other three 1181 pointings were additionally observed with one high resolution grating, with an exposure time of $2.6$~hours for the prism and each grating at each pointing. c: Four 1345 pointings were observed with both the prism and medium resolution gratings, another two pointings were observed with gratings only and there are additional two observed with the prism only. Five of six pointings were observed using the prism with an exposure time of $0.9$~hour at each pointing, the remaining one was observed with $1.7$~hours for prism.}
\label{tab:nirspec_obs}
\end{deluxetable*}

All the 2D NIRSpec spectra were reduced following the methods described in \citet{Topping2024a} using the standard \textit{JWST} data reduction pipeline\footnote{\url{https://github.com/spacetelescope/jwst}} \citep{Bushouse2024}. 
When performing the reduction, we also applied a wavelength-dependent slit loss correction for each spectrum assuming a point source, since the majority of objects in our sample are not significantly extended. 
We will discuss the slit loss in Section~\ref{sec:lya_evolution} when comparing against ground-based observations. 
For each object, we extract the 1D spectrum from the reduced 2D spectrum with a boxcar extraction. 
The extraction aperture is set to match the continuum or the emission line profile along the spatial direction, with a typical width of $\sim5$~pixels ($\sim0.5$~arcsec in spatial direction; see also \citealt{Tang2023,Chen2024}). 

\subsection{Sample Selection} \label{sec:sample}

Using the public NIRSpec dataset, we identify a sample of galaxies with spectroscopic redshifts at $z>6.5$. 
We determine the spectroscopic redshifts by visually inspecting the 2D spectra, searching for emission lines or the Ly$\alpha$ break. 
We identify $204$ sources at $z>6.5$ with emission line detections. 
Then we derive the accurate systemic redshifts ($z_{\rm sys}$) of these systems as follows. 
For each object, we simultaneously fit the available strong rest-frame optical emission lines (H$\gamma$, H$\beta$, [O~{\small III}], or H$\alpha$) with Gaussian profiles. 
We use the fitted line centers to compute redshifts, then average them weighted by emission line S/N to obtain a final redshift. 
One object (JADES-20096216) has strong optical lines ([O~{\small III}], Balmer lines) that are shifted out of the NIRSpec spectrum, so we derive the systemic redshift ($z=12.513$) based on its [Ne~{\small III}]~$\lambda3869$ detection (see also \citealt{Bunker2023b,Curtis-Lake2023,DEugenio2023}). 
Overall we measure the systemic redshifts of $204$ emission line sources at $6.533<z_{\rm sys}<12.513$. 

In addition to galaxies with emission line detections, we identify a subset of galaxies without emission line detections based on the presence of Ly$\alpha$ break. This is particularly important at $z\gtrsim9.5$ given the absence of strong rest-frame optical emission lines in NIRSpec spectra. 
We include $10$ sources across the different fields with Ly$\alpha$ break redshifts at $9.75<z<13.22$. Our measurements are consistent with literature redshifts for these sources \citep{ArrabalHaro2023a,ArrabalHaro2023b,Curtis-Lake2023,Fujimoto2023b}. 
In order to maximize the size of the sample and ensure we are not biased against sources with weak emission lines (e.g., extremely metal-poor galaxies), we also search for sources with Ly$\alpha$ breaks and no emission lines at lower redshifts ($6.5<z<9.5$). 
We find $4$ such objects with redshifts at $6.69<z<7.61$. 
In the following Section~\ref{sec:photometry}, we will examine whether the Ly$\alpha$ break identifications of these systems are valid by fitting their SEDs. 
In total the $z>6.5$ NIRSpec sample contains $218$ galaxies with redshifts at $6.53<z<13.22$. 

Our goal is to investigate Ly$\alpha$ properties of galaxies dominated by star formation, thus we remove systems hosting active galactic nuclei (AGNs) from the sample. 
Among the $218$ sources at $z>6.5$, we find $8$ objects showing very broad Balmer emission lines (H$\beta$ or H$\alpha$), with full width at half maximum (FWHM) $>1500$~km~s$^{-1}$ and narrow [O~{\small III}] emission lines (FWHM $<300$~km~s$^{-1}$). 
This strongly suggests that they are broad-line AGNs, and indeed these sources have previously been identified as AGNs in the literature \citep{Fujimoto2023b,Furtak2023a,Kocevski2024,Roberts-Borsani2024}. 
We remove these $8$ systems from our sample, leaving $210$ galaxies at $z>6.5$. 
There are $32$, $68$, $65$, and $45$ $z>6.5$ galaxies in the Abell 2744, EGS, GOODS-N, and GOODS-S fields, respectively (Table~\ref{tab:galaxy_num}). 
These $210$ galaxies consist of $196$ emission line galaxies and $14$ Ly$\alpha$ break systems without emission line detection. 
Among the $210$ galaxies, $188$ objects have low resolution prism spectra and $132$ have medium or high resolution grating spectra, with $110$ galaxies have both prism and grating spectra. 
In Table~\ref{tab:sources} we list the $210$ galaxies in our $z>6.5$ sample. 
This sample is $3-6\times$ larger than that used in earlier studies of Ly$\alpha$ emission with NIRSpec ($\sim30-70$ objects at $z>6.5$; e.g., \citealt{Jones2024,Nakane2024,Napolitano2024}).


\begin{deluxetable}{ccc}
\tablecaption{Number counts of $z>6.5$ galaxies identified from NIRSpec dataset and number counts of galaxies with Ly$\alpha$ detections in each field.}
\tablehead{
Field & \# of galaxies & \# of LAEs 
}
\startdata
Abell 2744 & $32$ & $4$ \\
EGS & $68$ & $15$ \\
GOODS-N & $65$ & $8$ \\
GOODS-S & $45$ & $6$ \\
\enddata
\end{deluxetable}
\label{tab:galaxy_num}

Next we search for Ly$\alpha$ emission lines of $z>6.5$ galaxies based on the systemic redshifts. 
We detect Ly$\alpha$ emission with S/N $>3$ in $33$ galaxies out of the total $210$ objects. 
In $17$ galaxies their Ly$\alpha$ emission lines are detected in grating spectra, and in $25$ galaxies the Ly$\alpha$ lines are detected in prism spectra, including $9$ galaxies with Ly$\alpha$ lines detected in both grating and prism spectra. 
The Ly$\alpha$ detections of $30$ of these $33$ Ly$\alpha$ emitting galaxies at $z>6.5$ have been reported in literature \citep{Bunker2023a,Fujimoto2023b,Tang2023,Tang2024a,Tang2024b,Chen2024,Jones2024,Napolitano2024,Saxena2024,Witstok2024b}. 
We have newly identified $3$ galaxies with Ly$\alpha$ detections (JADES-13041, JADES-14373, JADES-15423), which will be described in detail in Section~\ref{sec:new_lae}. 
We will derive the Ly$\alpha$ EWs, Ly$\alpha$ escape fractions, and Ly$\alpha$ velocity offsets of galaxies in our $z>6.5$ sample in Section~\ref{sec:measurements}. 

\subsection{Photometric Measurements} \label{sec:photometry}

We now characterize the physical properties (stellar populations, dust attenuation, ionizing properties) of galaxies in our $z>6.5$ NIRSpec sample by fitting their SEDs. 
We use the photometry extracted from the public \textit{JWST} Near Infrared Camera (NIRCam; \citealt{Rieke2005,Rieke2023}) imaging data when available. 
The NIRCam images were obtained from the observation programs introduced in Section~\ref{sec:spectra}. 
In all programs, we use deep imaging taken with six broad-band filters (F115W, F150W, F200W, F277W, F356W, and F444W) and one medium-band filter (F410M). 
For sources in JADES, we use additional images taken with F090W and F335M. 
We also use the F090W image taken from UNCOVER for sources in the Abell 2744 field. 
The NIRCam mosaics cover $197$ of the total $210$ galaxies in our $z>6.5$ sample. 
For $185$ of these $197$ galaxies, NIRCam images were taken with filters from F090W/F115W to F444W, fully covering their rest-frame UV to optical SEDs. 
For the remaining $12$ sources, NIRCam data were only taken with $1-3$ filters. 
The remaining $13$ of $210$ galaxies have not been observed with NIRCam, and we utilize the \textit{Hubble Space Telescope} (\textit{HST}) images from the Cosmic Assembly Near-infrared Deep Extragalactic Legacy Survey (CANDELS; \citealt{Grogin2011,Koekemoer2011}) dataset. 
We use the \textit{HST} images taken with the Wide Field Camera 3 (WFC3/IR) filters F125W and F160W, which covers the rest-frame UV wavelengths of $z>6.5$ galaxies. 
When fitting SEDs, we will focus on the $185$ galaxies with full (rest-frame UV to optical) NIRCam SEDs in our sample. 

We leverage the reduced \textit{JWST}/NIRCam images and the photometry catalogs released on the DAWN \textit{JWST} Archive\footnote{\url{https://dawn-cph.github.io/dja/}} (DJA). 
All the images were reduced with the grizli\footnote{\url{https://github.com/gbrammer/grizli}} code \citep{Brammer2023} in a homogeneous way. 
The process is briefly described in \citet{Valentino2023} and the full details will be presented in Brammer et al. (in prep.). 
The source extraction and photometry measurement on the reduced image were performed by the DJA team, using the Python version of \texttt{Source Extractor} \citep{Bertin1996} \texttt{sep} \citep{Barbary2016}. 
We use the photometry measured in $0.5$~arcsec diameter circular apertures and corrected to the ``total'' fluxes within elliptical Kron \citep{Kron1980} apertures. 
For galaxies in the Abell 2744 lensing cluster field (i.e., from GLASS and UNCOVER observations), we apply the magnification corrections\footnote{\url{https://jwst-uncover.github.io/DR4.html}} \citep{Price2024} derived from the updated lensing models in \citet{Furtak2023b} to their photometry. 
For \textit{HST} imaging we use the CANDELS photometry catalogs generated by \citet{Guo2013} and \citet{Stefanon2017}. 

\begin{figure}
\includegraphics[width=\linewidth]{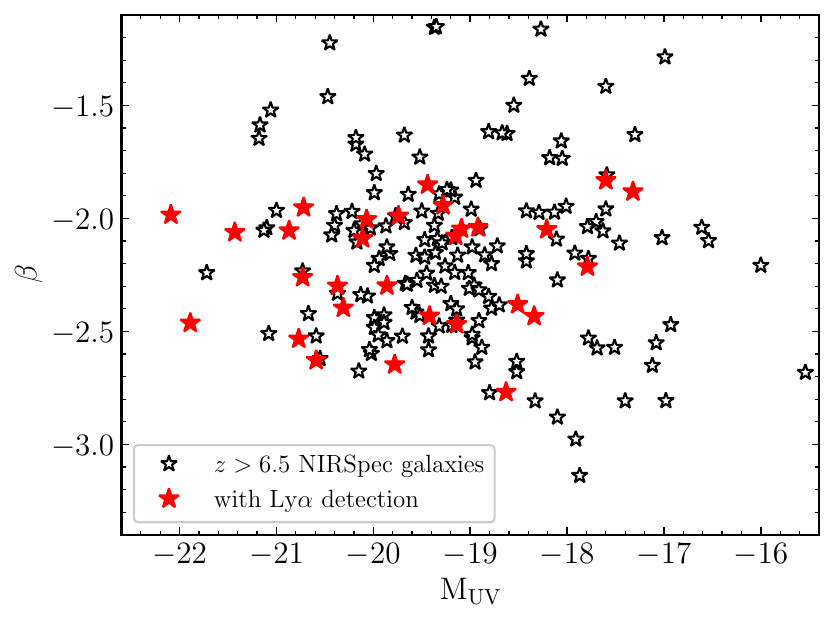}
\caption{Absolute UV magnitude (M$_{\rm UV}$) and UV continuum slope ($\beta$) of $z>6.5$ galaxies identified from the public NIRSpec observations used in this study. We show the $185$ galaxies with both M$_{\rm UV}$ and UV slope measurements (open black stars). We mark those with Ly$\alpha$ detections as red filled stars.}
\label{fig:muv_beta}
\end{figure}

For each galaxy in our $z>6.5$ NIRSpec sample, we cross-match its coordinate with the photometry catalog and identify the best-matched source by visually inspecting the image. 
The absolute UV magnitudes (M$_{\rm UV}$) of the $210$ galaxies in our $z>6.5$ sample range from $-22.1$ to $-15.5$, with a median M$_{\rm UV}=-19.4$ (Figure~\ref{fig:muv_beta}). 
Here the M$_{\rm UV}$ of the $32$ systems in the Abell 2744 field have been corrected for gravitational lensing with the \citet{Furtak2023b} model \citep{Price2024}, with a median magnification correction factor $\mu=1.9$. 
We derive the UV continuum slopes for the $185$ galaxies with NIRCam SEDs by fitting a power law ($f_{\lambda}\propto\lambda^{\beta}$) to the broad-band photometry at rest-frame wavelength $1250-2600$~\AA\ \citep{Calzetti1994}. 
The UV slopes of these $185$ galaxies at $z>6.5$ are generally blue (median $\beta=-2.2$; Figure~\ref{fig:muv_beta}), suggesting relatively low dust attenuation, consistent with typical galaxies at these redshifts \citep[e.g.,][]{Finkelstein2012,Bouwens2014,Cullen2023,Topping2024b}. 


We fit the NIRCam SEDs of the $185$ galaxies at $z>6.5$ using the Bayesian galaxy SED modelling and interpreting tool BayEsian Analysis of GaLaxy sEds (\texttt{BEAGLE}, version 0.29.2; \citealt{Chevallard2016}). 
The \texttt{BEAGLE} tool utilizes the latest version of the \citet{Bruzual2003} stellar population synthesis models and the \citet{Gutkin2016} photoionization models of star-forming galaxies with the \texttt{CLOUDY} code \citep{Ferland2013}. 
For each emission line galaxy, we fix the redshift to its systemic redshift measured from the NIRSpec spectrum (Section~\ref{sec:sample}). 
For Ly$\alpha$ break galaxies with no emission lines, we fit their redshifts in the range $0<z<15$ with a uniform prior. 
We assume a constant star formation history (CSFH), allowing the galaxy age to vary between $1$~Myr and the age of the Universe at the given redshift with a log-uniform prior. 
We also assume the \citet{Chabrier2003} initial mass function (IMF) with stellar mass range of $0.1-300\ M_{\odot}$. 
The metallicity is set to vary in the range $-2.2\le\log{(Z/Z_{\odot})}\le0.25$ ($Z_{\odot}=0.01524$; \citealt{Caffau2011}) and the dust-to-metal mass ratio ($\xi_{\rm d}$) spans the range $\xi_{\rm d}=0.1-0.5$. 
The ionisation parameter $U$ is adjusted in the range $-4.0\le\log{U}\le-1.0$. 
We adopt log-uniform priors for metallicity and ionization parameter, and a uniform prior for dust-to-metal mass ratio. 
For the dust attenuation, we apply the Small Magellanic Cloud (SMC) extinction curve \citep{Pei1992}, allowing the $V$-band optical depth $\tau_{V}$ to vary between $0.001$ and $5$ with a log-uniform prior. 
Finally, we apply the prescription of \citet{Inoue2014} to account for the absorption of IGM. 
When fitting SEDs, we exclude the fluxes in filters blueward of Ly$\alpha$ to avoid introducing the uncertain contribution from Lyman series emission and absorption. 

From \texttt{BEAGLE} SED fitting, we derive the median values and the marginalized $68\%$ credible intervals from the posterior probability distributions for the fitted parameters. 
For the $185$ galaxies at $z>6.5$ with NIRCam SEDs, \texttt{BEAGLE} models indicate that their SEDs are dominated by young stellar populations, with CSFH ages spanning from $1.4$~Myr to $575$~Myr (median age $=28$~Myr). 
The specific star formation rates (sSFRs) are large, with sSFR $=2-724$~Gyr$^{-1}$ (median sSFR $=36$~Gyr$^{-1}$) assuming CSFH. 
These suggest that many of our $z>6.5$ galaxies have recently experienced substantial bursts or upturns in their star formation histories, as is typical at these redshifts \citep[e.g.,][]{Endsley2021,Endsley2023,Boyett2024}.

The inferred dust attenuation of our $z>6.5$ galaxies is low, with a median rest-frame $V$-band optical depth only $\tau_{V}=0.04$. 
This is consistent with the blue UV slopes of our systems. 
The CSFH stellar masses of these $185$ galaxies are relatively low, ranging from $M_{\star}=2.6\times10^6\ M_{\odot}$ to $1.7\times10^9\ M_{\odot}$ (median $M_{\star}=4.9\times10^7\ M_{\odot}$). 
Here we note that the stellar masses derived from CSFH models correspond to the young stellar populations dominating the rest-frame UV to optical light. 
If older stellar populations exist in these galaxies (which could be easily outshined by the light from young stars), the stellar masses may increase up to an order of magnitude \citep[e.g.,][]{Tang2022,Tacchella2023,Whitler2023}. 
However, this effect does not impact the main results of this paper, as we do not discuss any trends with stellar mass. 

The NIRCam SEDs exhibit flux excesses in filters contaminated by [O~{\small III}] and H$\beta$, allowing useful constraints to be placed on the [O~{\small III}]+H$\beta$ EWs of galaxies in 
our sample. We find the rest-frame [O~{\small III}]+H$\beta$ EWs span from $126$~\AA\ to $5463$~\AA\ for the $185$ galaxies with \texttt{BEAGLE} models, with a median value of $735$~\AA. 
This median [O~{\small III}]+H$\beta$ EW is comparable to the average [O~{\small III}]+H$\beta$ EW of galaxies at $z>6$ ($\simeq700-800$~\AA; e.g., \citealt{Endsley2023,Endsley2024}). 
The strong emission lines are expected given the young stellar ages inferred from the \texttt{BEAGLE} models for our sources \citep[e.g.,][]{Chevallard2018,Tang2019,Tang2023}. 

The NIRCam SEDs also allow us to constrain the hydrogen ionizing photon production efficiencies ($\xi_{\rm ion}$). 
Here we adopt the most commonly used definition of $\xi_{\rm ion}$ in literature: the hydrogen ionizing photon production rate per unit intrinsic UV luminosity density at rest-frame $1500$~\AA\ ($L_{\rm UV}$), where $L_{\rm UV}$ includes both stellar and nebular continuum and is corrected for dust attenuation (see \citealt{Chevallard2018} for various definitions of $\xi_{\rm ion}$ used in the literature). 
The \texttt{BEAGLE} models indicate that our galaxies have large ionizing photon production efficiencies, ranging from $\xi_{\rm ion}=10^{25.4}$~erg$^{-1}$~Hz to $10^{25.9}$~erg$^{-1}$~Hz with a median $\xi_{\rm ion}=10^{25.5}$~erg$^{-1}$~Hz. 
These are comparable to the reionization-era populations (median $\xi_{\rm ion}\simeq10^{25.5-25.7}$~erg$^{-1}$~Hz; e.g., \citealt{Tang2023,Endsley2024}). 
These results suggest that many of our $z>6.5$ galaxies have hard ionizing spectra. 

Finally, we examine the Ly$\alpha$ break identifications of the $14$ galaxies with no emission lines using the SED fitting results. 
Comparing to the redshifts derived from Ly$\alpha$ breaks in NIRSpec spectra, the posterior median redshifts inferred from \texttt{BEAGLE} models are consistent within $\Delta z<0.3$. 
For the four galaxies at $6.5<z<9.5$, we consider whether the emission line fluxes predicted from \texttt{BEAGLE} models are consistent with the non-detections in the NIRSpec spectra. 
For [O~{\small III}]~$\lambda5007$ (which is often the brightest optical emission line), the posterior median fluxes inferred from \texttt{BEAGLE} models for these four systems are $0.5-4\times10^{-19}$~erg~s~cm$^{-2}$. 
These are in agreement with the $3\sigma$ upper limits of [O~{\small III}]~$\lambda5007$ fluxes measured from NIRSpec spectra ($<2-5\times10^{-19}$~erg~s~cm$^{-2}$). 
Therefore, we conclude that the Ly$\alpha$ break identifications of the $14$ galaxies without emission line detections from NIRSpec spectra are consistent with their NIRCam SEDs. 

\subsection{New Ly$\alpha$ Emitting Galaxies at $z>6.5$} \label{sec:new_lae}


\begin{figure*}
\includegraphics[width=\linewidth]{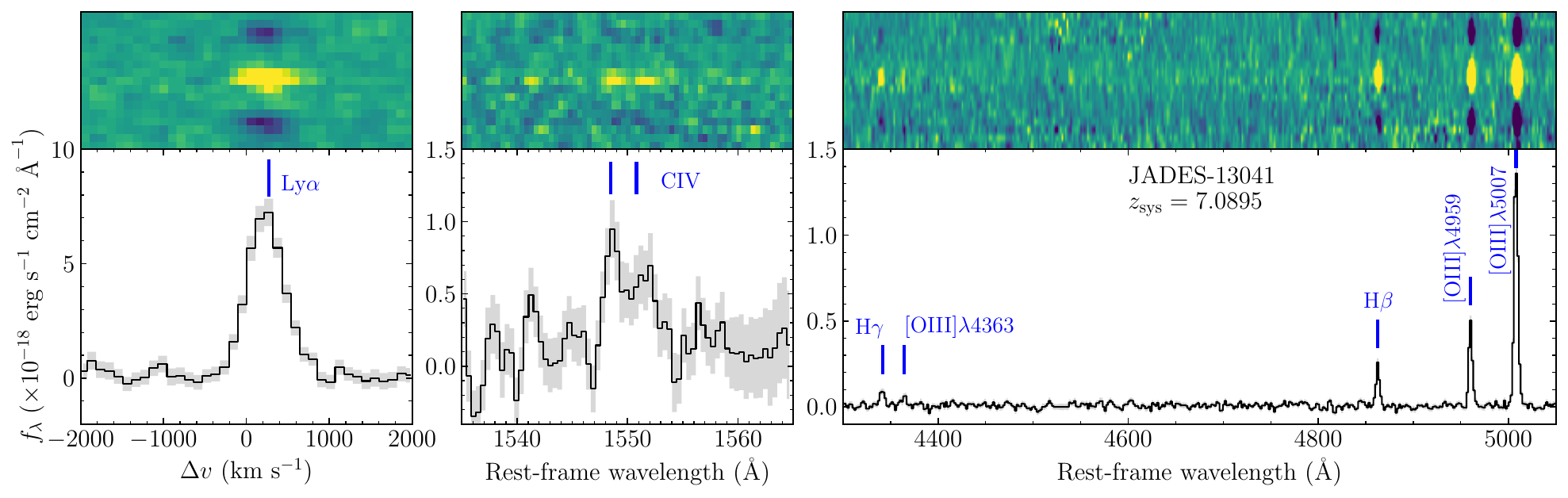}
\caption{\textit{JWST}/NIRSpec 2D and 1D medium resolution grating spectra of JADES-13041. We show the Ly$\alpha$ velocity profile (left panel), the C~{\scriptsize IV} emission lines (middle panel), and strong rest-frame optical emission lines (H$\gamma$, H$\beta$, [O~{\scriptsize III}]; right panel).}
\label{fig:JADES-13041_spec}
\end{figure*}

Our analysis has revealed three galaxies in the public datasets with newly identified Ly$\alpha$ emission detections. 
We present the spectroscopic detections and properties of these galaxies in this subsection. 

The strongest Ly$\alpha$ emitter in the new sample is JADES-13041 at $z=7.1$, identified in the 1181 observations of GOODS-N. 
This galaxy is situated at the same redshift as several other Ly$\alpha$ emitters in this field (\citealt{Tang2024b}; see also Section~\ref{sec:lya_association}). 
In Figure~\ref{fig:JADES-13041_spec}, we show the medium resolution grating spectra of JADES-13041. 
Strong rest-frame optical emission lines (H$\beta$, [O~{\small III}]) are clearly seen in its G395M/F290LP spectrum. 
By simultaneously fitting the strong optical lines with Gaussian profiles, we derive a systemic redshift of $z_{\rm sys}=7.0895$. 
We detect the Ly$\alpha$ emission line of JADES-13041 in its G140M/F070LP spectrum, with $z_{{\rm Ly}\alpha}=7.0958$. 
This indicates a Ly$\alpha$ velocity offset of $\Delta v_{{\rm Ly}\alpha}=234\pm53$~km~s$^{-1}$. 
We measure a Ly$\alpha$ flux of $F_{{\rm Ly}\alpha}=1.61\pm0.09\times10^{-17}$~erg~s$^{-1}$~cm$^{-2}$ and EW $=143\pm8$~\AA. 
The Ly$\alpha$ EW of JADES-13041 places it in the top $3\%$ of the Ly$\alpha$ EW distribution at $z\simeq7$ (see Section~\ref{sec:lya_ew}), indicating its strong Ly$\alpha$ that is atypical in the reionization era. 

We also constrain the Ly$\alpha$ escape fraction of JADES-13041 using the H$\beta$ emission line. 
We measure a H$\gamma$/H$\beta$ ratio of $0.49\pm0.03$. 
Comparing to the intrinsic H$\gamma$/H$\beta$ ratio expected in case B recombination ($0.47$), this indicates negligible dust attenuation to the nebular emission. 
Assuming case B recombination with $T_{\rm e}=10^4$~K, we derive a large Ly$\alpha$ escape fraction of $f_{{\rm esc,Ly}\alpha}=0.57\pm0.12$.
We note that different recombination assumptions may yield different escape fractions \citep[e.g.,][]{Chen2024,McClymont2024,Scarlata2024,Tang2024a,Yanagisawa2024}, but the net conclusion that JADES-13041 transmits a large fraction of its Ly$\alpha$ emission will not change. 
The fact that such a strong Ly$\alpha$ emitter is located at the same redshift as other Ly$\alpha$ emitters in GOODS-N may suggest the presence of an ionized bubble. 
We will come back to this in Section~\ref{sec:lya_association}, describing whether there is evidence for an overdensity of galaxies at this redshift.

The spectrum of JADES-13041 reveals high ionization nebular C~{\small IV}~$\lambda\lambda1548,1551$ emission in the G140M/F070LP spectrum. 
We measure a C~{\small IV} doublet EW $=21\pm5$~\AA, consistent with the intense C~{\small IV} emission seen in a subset of reionization-era galaxies with metal poor gas \citep[e.g.,][]{Stark2015,Castellano2024,Tang2024b,Topping2024a,Witstok2024b}. 
The line ratio of the C~{\small IV} doublet components ($1548/1551=1.6$) is close to the intrinsic value ($1548/1551=2$; e.g., \citealt{Flower1979}), consistent with minimal scattering of the resonant line photons. 
The detection of such strong C~{\small IV} emission indicates a hard radiation field in JADES-13041. 
The hard radiation field may efficiently ionize the surrounding IGM, increasing the ionizing fraction \citep{Mason2020}. 
This may allow Ly$\alpha$ photons near the systemic redshift to escape (left panel of Figure~\ref{fig:JADES-13041_spec}), boosting the transmission of Ly$\alpha$ emission for JADES-13041. 


\begin{figure}
\includegraphics[width=\linewidth]{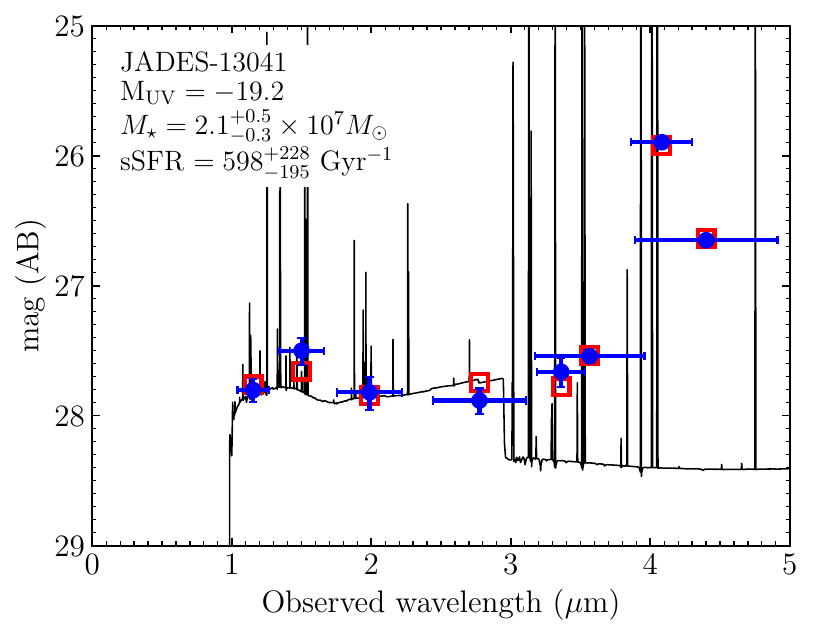}
\caption{\textit{JWST}/NIRCam SED of JADES-13041. Observed NIRCam photometry is shown by blue circles. The \texttt{BEAGLE} model spectrum is shown by the black line and the synthetic photometry is presented by open red squares.}
\label{fig:JADES-13041_sed}
\end{figure}

We show the NIRCam photometry of JADES-13041 and the best-fit \texttt{BEAGLE} model (Section~\ref{sec:photometry}) in Figure~\ref{fig:JADES-13041_sed}. 
JADES-13041 has an absolute UV magnitude of $M_{\rm UV}=-19.2$. 
Its UV slope is blue ($\beta=-2.1$), consistent with the negligible dust attenuation inferred from H$\gamma$/H$\beta$ ratio. 
The SED fitting results indicate a relatively low stellar mass ($M_{\star}=2.1^{+0.5}_{-0.3}\times10^7\ M_{\odot}$ assuming CSFH). 
\texttt{BEAGLE} models also demonstrate that the rest-frame UV to optical light of this galaxy is dominated by very young stellar populations (CSFH age $=1.7^{+0.8}_{-0.5}$~Myr), as expected for galaxies undergoing recent upturns in star formation. 


\begin{figure}
\includegraphics[width=\linewidth]{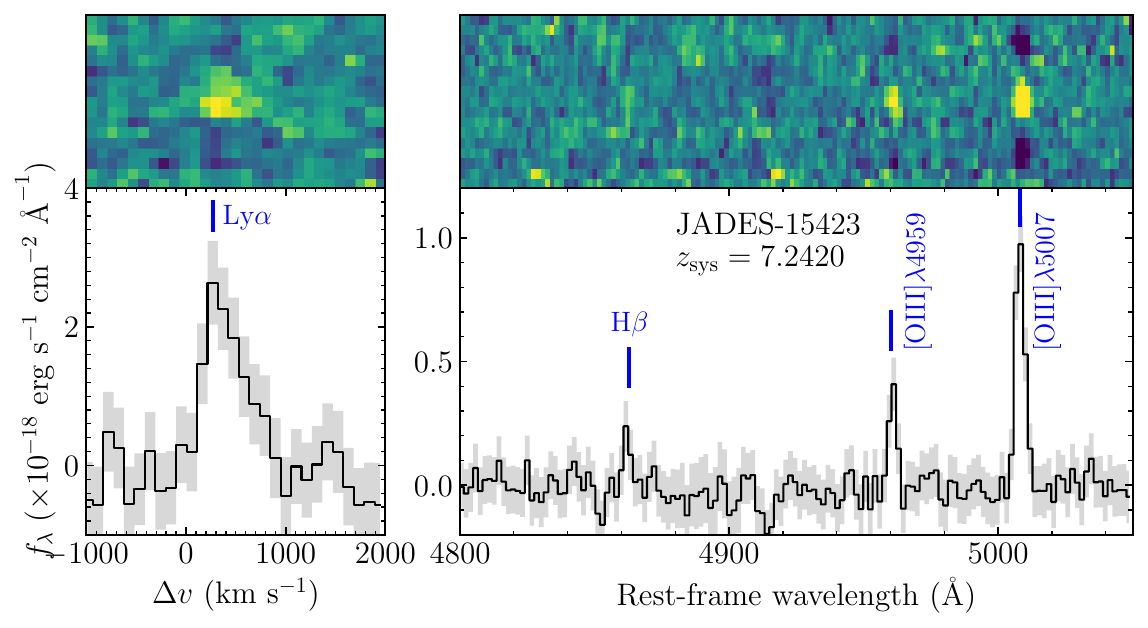}
\includegraphics[width=\linewidth]{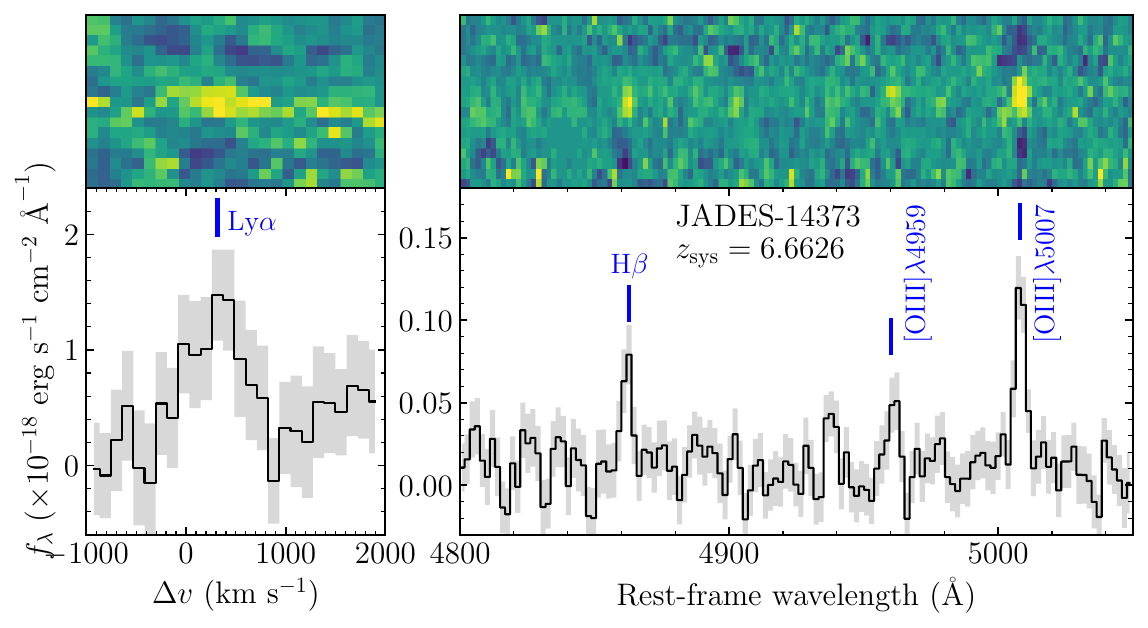}
\caption{2D and 1D NIRSpec medium resolution grating spectra of JADES-15423 (top) and JADES-14373 (bottom). The left panels show the Ly$\alpha$ velocity profiles. The right panels show H$\beta$ and [O~{\scriptsize III}] emission lines.}
\label{fig:new_grating}
\end{figure}

The two other new Ly$\alpha$ detections are JADES-15423 (identified in the 1180 observations in GOODS-S) and JADES-14373 (identified in the 1181 observations in GOODS-N), both showing more moderate strength lines (EW $=25-50$~\AA) in the medium resolution gratings. 
The spectrum of JADES-15423 (M$_{\rm UV}=-20.1$) is shown in the top panel of Figure~\ref{fig:new_grating}. 
We find H$\gamma$, H$\beta$, and [O~{\small III}] emission lines in the G395M/F290LP spectrum.
By fitting these optical lines with Gaussian profiles, we derive a systemic redshift of $z_{\rm sys}=7.2420$. 
In its G140M/F070LP spectrum, we detect the Ly$\alpha$ emission with $z_{{\rm Ly}\alpha}=7.2494$, indicating a Ly$\alpha$ velocity offset of $\Delta v_{{\rm Ly}\alpha}=269\pm52$~km~s$^{-1}$. 
We measure a Ly$\alpha$ flux of $F_{{\rm Ly}\alpha}=4.77\pm0.66\times10^{-18}$~erg~s$^{-1}$~cm$^{-2}$ and EW $=26\pm3$~\AA. 
Then we estimate the dust attenuation from the Balmer decrement and hence calculate the Ly$\alpha$ escape fraction for JADES-15423. 
We measure a H$\gamma$/H$\beta$ ratio $=0.55\pm0.11$, indicating a negligible dust attenuation to the nebular emission by comparing with the intrinsic H$\gamma$/H$\beta$ ratio expected from case B recombination ($0.47$). 
Assuming case B recombination, we derive a Ly$\alpha$ escape fraction of $f_{{\rm esc,Ly}\alpha}=0.24\pm0.03$ for JADES-15423. 

For JADES-14373 (M$_{\rm UV}=-18.9$), we detect the H$\beta$, [O~{\small III}], and H$\alpha$ emission lines (bottom panel of Figure~\ref{fig:new_grating}). 
We fit the optical lines with Gaussians and derive a systemic redshift of $z_{\rm sys}=6.6626$. 
We detect Ly$\alpha$ at $z_{{\rm Ly}\alpha}=6.6704$, indicating a velocity offset of $\Delta v_{{\rm Ly}\alpha}=305\pm56$~km~s$^{-1}$. 
The Ly$\alpha$ flux is $F_{{\rm Ly}\alpha}=2.97\pm0.91\times10^{-18}$~erg~s$^{-1}$~cm$^{-2}$ and the resulting EW is $=49\pm15$~\AA. 
We infer the nebular dust attenuation of JADES-15423 from the H$\alpha$/H$\beta$ ratio ($=2.69\pm0.46$). 
Comparing with the intrinsic value expected from case B recombination (H$\alpha$/H$\beta=2.87$), this suggests negligible dust attenuation to nebular gas. 
We then derive a Ly$\alpha$ escape fraction of $f_{{\rm esc,Ly}\alpha}=0.38\pm0.12$ for JADES-14373 assuming case B recombination. 

\subsection{Spectroscopic Measurements} \label{sec:measurements}


\begin{figure}
\includegraphics[width=\linewidth]{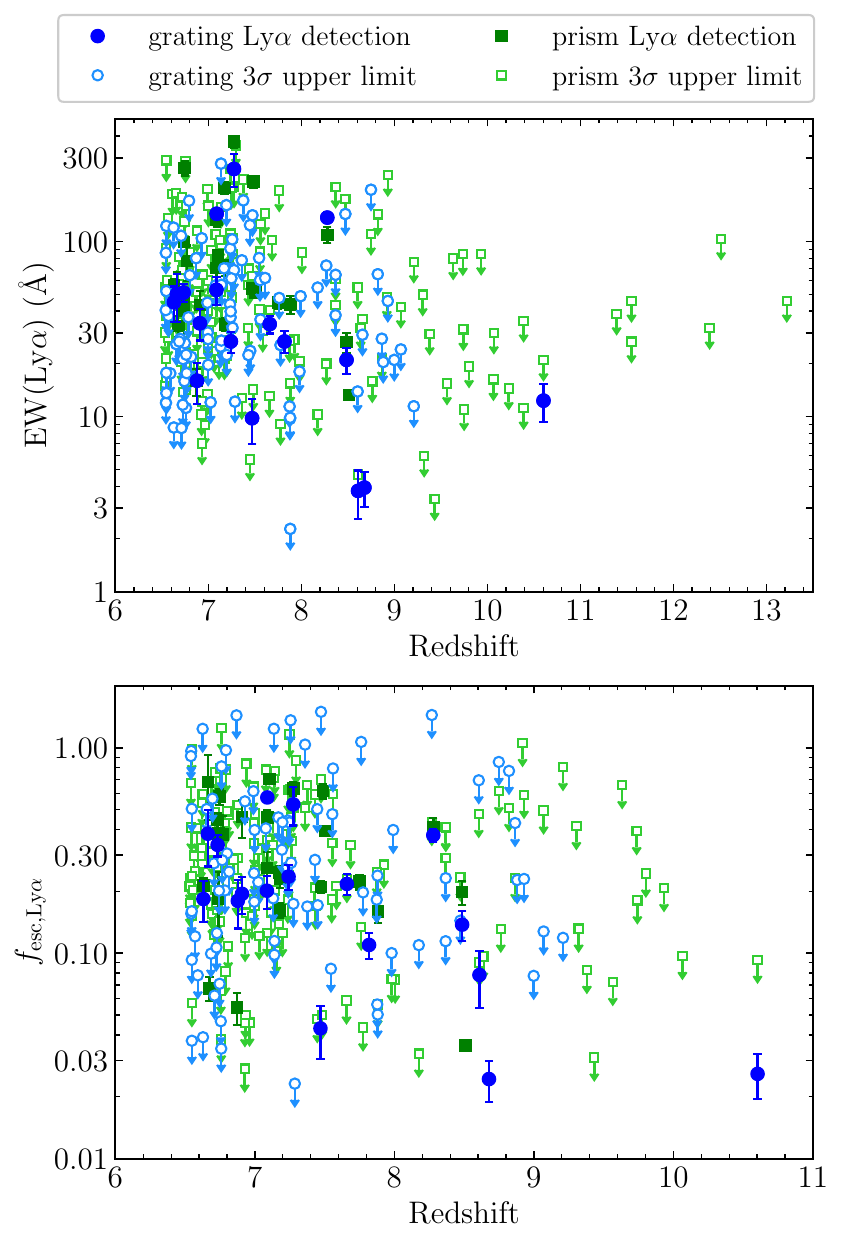}
\caption{Ly$\alpha$ EWs (top panel) and Ly$\alpha$ escape fractions (bottom panel) versus redshift of galaxies at $z>6.5$ in the public NIRSpec sample. We show the Ly$\alpha$ EWs and Ly$\alpha$ escape fractions measured from grating spectra (blue circles) and prism spectra (green squares) separately. Sources with Ly$\alpha$ detections are shown by filled symbols, and we put $3\sigma$ upper limits for Ly$\alpha$ non-detections as open symbols.}
\label{fig:lyaew_fesc_z}
\end{figure}

In this subsection, we measure the emission line properties of the $210$ galaxies in our $z>6.5$ sample from the NIRSpec spectra. 
We first measure the Balmer emission line (H$\delta$, H$\gamma$, H$\beta$, or H$\alpha$) fluxes and derive the nebular dust attenuation using Balmer decrements. 
Then we derive the fluxes of Ly$\alpha$ emission lines as well as the Ly$\alpha$ EWs, Ly$\alpha$ escape fractions, and Ly$\alpha$ velocity offsets. 
The emission line measurements are performed for both the NIRSpec prism and grating spectra. 

We derive the Balmer emission line fluxes by fitting the line profiles with Gaussians for the $196$ emission line galaxies in our sample (except for JADES-20096216 whose Balmer lines are not visible, see Section~\ref{sec:sample}). 
To evaluate the uncertainties of measured line fluxes, we resample the flux densities of each spectrum $1000$ times by taking the observed flux densities as mean values and the errors as standard deviations. 
Then we derive the line fluxes from the resampled spectra of each object and take the standard deviation as the uncertainty. 

Utilizing the measured Balmer emission line fluxes, we estimate the dust attenuation to nebular emission from Balmer decrement. 
We use the observed H$\alpha$/H$\beta$ flux ratios for galaxies at $z\simeq6.5-7.0$ where both emission lines are visible in NIRSpec spectra, the H$\gamma$/H$\beta$ ratios for those at $z\simeq7.0-9.5$ where H$\alpha$ is not visible, and the H$\delta$/H$\gamma$ ratios for those at $z\simeq9.5-10.6$ where neither H$\alpha$ nor H$\beta$ is visible. 
Assuming case B recombination and a gas temperature $T_{\rm e}=10^4$~K, we expect the intrinsic H$\alpha$/H$\beta$, H$\gamma$/H$\beta$, and H$\delta$/H$\gamma$ ratios are $2.87$, $0.47$, and $0.55$, respectively \citep{Osterbrock2006}. 
We compare the observed Balmer decrements to the intrinsic values to derive the dust extinction assuming the \citet{Cardelli1989} curve. 
We infer a median $E(B-V)=0.01$, suggesting negligible dust attenuation to the nebular emission of our $z>6.5$ galaxies. 
This result is consistent with the low dust attenuation inferred from \texttt{BEAGLE} SED fitting (Section~\ref{sec:photometry}). 

We now compute the Ly$\alpha$ fluxes and EWs from grating spectra. 
For each object with grating spectrum, we calculate the underlying continuum flux density by fitting the spectrum at rest-frame $1300$~\AA\ to $1400$~\AA\ with a flat continuum (in $f_\nu$) and extrapolating the fitted continuum to the Ly$\alpha$ line center (determined using the systemic redshift). 
We do not use the spectrum at rest-frame $<1300$~\AA\ for continuum fitting to avoid potential damped Ly$\alpha$ absorption. 
For each galaxy with a Ly$\alpha$ detection in grating spectrum ($17$ galaxies), we derive its Ly$\alpha$ flux by fitting the continuum-subtracted line profile with an asymmetric Gaussian function to account for the impact of the IGM to the blue side of the Ly$\alpha$ line. 
The uncertainty of the Ly$\alpha$ flux is estimated using the same resampling methods introduced above. 
For those which Ly$\alpha$ is undetected in grating spectra ($85$ galaxies), we put $3\sigma$ upper limits to the Ly$\alpha$ fluxes by directly integrating the error spectra over $\pm1000$~km~s$^{-1}$ within the systemic redshift in quadrature. 
This window ensures we capture the entire Ly$\alpha$ profile of a $z\gtrsim6$ galaxy \citep[e.g.,][]{Tang2024a,Saxena2024}. 
Then the Ly$\alpha$ EWs (upper limits) are computed as the ratio between the measured Ly$\alpha$ line fluxes (upper limits) and the underlying continuum flux densities. 
For the $17$ galaxies with Ly$\alpha$ detections in grating spectra, we derive the $16{\rm th}-50{\rm th}-84{\rm th}$ percentile of Ly$\alpha$ EW $=11$~\AA, $34$~\AA, and $90$~\AA. 
For the $85$ galaxies with Ly$\alpha$ non-detections, the median $3\sigma$ upper limit of their Ly$\alpha$ EWs derived from grating spectra is $<36$~\AA. 

We apply the following methods to calculate the Ly$\alpha$ fluxes and EWs from prism spectra. 
Due to the low resolution of the blue end of NIRSpec prism spectra ($R\sim30$), the Ly$\alpha$ emission of reionization-era galaxies will spread pixels at both sides of the Ly$\alpha$ break \citep[e.g.,][]{Chen2024,Jones2024,Keating2024,Napolitano2024}. 
To estimate the underlying continuum flux density from prism spectrum, we fit the continuum with a power law at rest-frame $1300$~\AA\ to $1600$~\AA\ and a Heaviside function at the Ly$\alpha$ break \citep[e.g.,][]{Jones2024} for each object. 
For galaxies with Ly$\alpha$ detections in prism spectra ($25$ galaxies), we calculate the Ly$\alpha$ line fluxes by directly integrating the continuum-subtracted line profile over rest-frame $\simeq1170-1270$~\AA\ (corresponding to $\simeq6$~pixels centered on the Ly$\alpha$ peak in prism spectrum). 
We derive the $16{\rm th}-50{\rm th}-84{\rm th}$ percentile of Ly$\alpha$ EW $=34$~\AA, $54$~\AA, and $145$~\AA\ for these $25$ galaxies. 
For the $9$ galaxies with Ly$\alpha$ detections in both prism and grating spectra, we find that their Ly$\alpha$ fluxes and EWs derived from both spectra are generally consistent, with a standard deviation of the difference between both EWs $\sim0.09$~dex. 
For galaxies in which Ly$\alpha$ is undetected in prism spectra ($159$ galaxies), we derive the $3\sigma$ upper limits of Ly$\alpha$ line fluxes by integrating the error spectra over the same wavelength range in quadrature. 
The median $3\sigma$ upper limit of the Ly$\alpha$ EWs of these $159$ galaxies is $<44$~\AA. 
The Ly$\alpha$ EWs of our galaxies are shown in the top panel of Figure~\ref{fig:lyaew_fesc_z}. 

We next calculate the Ly$\alpha$ escape fractions using the Ly$\alpha$ flux constraints and Balmer emission line fluxes. 
We define the Ly$\alpha$ escape fraction ($f_{{\rm esc,Ly}\alpha}$) as the ratio between the observed Ly$\alpha$ flux ($F^{\rm obs}_{{\rm Ly}\alpha}$) and the intrinsic Ly$\alpha$ flux ($F^{\rm int}_{{\rm Ly}\alpha}$). 
The intrinsic Ly$\alpha$ flux is derived from the dust corrected H$\alpha$ flux ($F^{\rm corr}_{{\rm H}\alpha}$; at $z\simeq6.5-7.0$), H$\beta$ flux ($F^{\rm corr}_{{\rm H}\beta}$; at $z\simeq7.0-9.5$), or H$\gamma$ flux ($F^{\rm corr}_{{\rm H}\gamma}$; at $z\simeq9.5-10.6$), where the dust correction is estimated from the Balmer decrement measurement. 
Assuming case B recombination and $T_{\rm e}=10^4$~K, we calculate the intrinsic Ly$\alpha$ flux as $F^{\rm int}_{{\rm Ly}\alpha}=8.7\times F^{\rm corr}_{{\rm H}\alpha}$, $F^{\rm int}_{{\rm Ly}\alpha}=25.0\times F^{\rm corr}_{{\rm H}\beta}$, or $F^{\rm int}_{{\rm Ly}\alpha}=53.6\times F^{\rm corr}_{{\rm H}\gamma}$ \citep[e.g.,][]{Osterbrock2006,Hayes2015,Henry2015}. 
For the $17$ ($25$) galaxies with both Ly$\alpha$ and Balmer emission line detections in grating (prism) spectra, we derive the $16{\rm th}-50{\rm th}-84{\rm th}$ percentile of $f_{{\rm esc,Ly}\alpha}=0.06$, $0.19$, and $0.38$ ($f_{{\rm esc,Ly}\alpha}=0.16$, $0.26$, and $0.59$). 
We put $3\sigma$ upper limits of $f_{{\rm esc,Ly}\alpha}$ for sources which Ly$\alpha$ is undetected but Balmer emission lines are detected. 
The median $3\sigma$ upper limit is $<0.24$ measured from grating spectra ($82$ objects) and $<0.25$ from prism spectra ($145$ objects). 
In the bottom panel of Figure~\ref{fig:lyaew_fesc_z} we present the Ly$\alpha$ escape fractions of galaxies in our $z>6.5$ sample. 

Finally, we derive the Ly$\alpha$ velocity offsets ($\Delta v_{{\rm Ly}\alpha}$) for a subset of galaxies with Ly$\alpha$ detections at $z>6.5$. 
Because the derivation of Ly$\alpha$ velocity offset requires precise measurements of redshift, we only consider objects with systemic redshifts derived from multiple emission lines measured in the medium or high resolution grating spectra. 
There are $17$ galaxies with Ly$\alpha$ detections and systemic redshift measurements from grating spectra at $z>6.5$. 
We derive their Ly$\alpha$ redshifts ($z_{{\rm Ly}\alpha}$) by fitting the line profiles with asymmetric Gaussian functions and using the fitted line centers. 
The Ly$\alpha$ velocity offset is computed as $\Delta v_{{\rm Ly}\alpha}=c(z_{{\rm Ly}\alpha}-z_{\rm sys})/(1+z_{\rm sys})$, where $c$ is the speed of light. 
For those $17$ galaxies we derive the $16{\rm th}-50{\rm th}-84{\rm th}$ percentile of Ly$\alpha$ velocity offsets $=168$~km~s$^{-1}$, $277$~km~s$^{-1}$, and $447$~km~s$^{-1}$. 
The uncertainties of Ly$\alpha$ velocity offsets are estimated by resampling the flux densities $1000$ times and taking the standard deviation of Ly$\alpha$ velocity offsets derived from the resampled spectra. 
The Ly$\alpha$ velocity offset results were initially presented in \citet{Tang2024b}. 
In Table~\ref{tab:sources} we list the Ly$\alpha$ properties of galaxies in our $z>6.5$ NIRSpec sample.


\section{The Evolution of L\lowercase{y}$\alpha$ at $\lowercase{z}\gtrsim6.5$} \label{sec:lya_evolution}

In this section, we derive the distributions of Ly$\alpha$ EWs and Ly$\alpha$ escape fractions at $z>6.5$. 
We introduce the methodology for deriving Ly$\alpha$ property distributions in Section~\ref{sec:lya_method}. 
Then we present the distributions of Ly$\alpha$ EW in Section~\ref{sec:lya_ew} and Ly$\alpha$ escape fraction in Section~\ref{sec:lya_fesc}. 
Using these distributions we quantify the fractions of galaxies presenting large Ly$\alpha$ EWs and large Ly$\alpha$ escape fractions at $z>6.5$ and explore the evolution of such Ly$\alpha$ emitter fractions in the reionization era. 
Finally, we discuss the field-to-field variations in Ly$\alpha$ in Section~\ref{sec:lya_variation}. 

\subsection{Methodology} \label{sec:lya_method}

We establish the Ly$\alpha$ EW and Ly$\alpha$ escape fraction distributions at $z>6.5$ using a Bayesian approach \citep[e.g.,][]{Schenker2014,Endsley2021,Boyett2022,Chen2024} following the methodology described in \citet{Tang2024a}. 
To be consistent with previous studies of Ly$\alpha$ property distributions at $z\gtrsim5$ \citep[e.g.,][]{Schenker2014,Endsley2021,Chen2024,Tang2024a}, we assume a log-normal distribution of Ly$\alpha$ EW or Ly$\alpha$ escape fraction for our sample. 
We note that fitting our data with a different distribution model (an exponentially declining function; \citealt{Dijkstra2011}) does not impact the main results (i.e., the fraction of galaxies showing large Ly$\alpha$ EWs or large $f_{{\rm esc,Ly}\alpha}$) significantly. 

We briefly summarize the methodology for constructing Ly$\alpha$ EW and Ly$\alpha$ escape fraction distributions. 
We model the distributions with a set of parameters $\theta=[\mu,\sigma]$, where $\mu$ and $\sigma$ are the mean and the standard deviation of a log-normal distribution. 
For a Ly$\alpha$ EW distribution, we assume uniform priors for the parameters: $\mu=0-6$ (corresponding to median Ly$\alpha$ EW $=1-400$~\AA) and $\sigma=0.01-3$ \citep[e.g.,][]{Schenker2014,Endsley2021}. 
For a Ly$\alpha$ escape fraction distribution, we adopt a uniform prior for $\mu$ ($=-9$ to $0$, or median $f_{{\rm esc,Ly}\alpha}=0.0001-1$) and a Gaussian prior for $\sigma$ (mean $=0.6$, standard deviation $=0.3$; \citealt{Chen2024}). 
Using Bayes' theorem, we derive the posterior probability distributions of the model parameters as: 
\begin{equation*}
p(\theta|{\rm obs}) \propto p(\theta) \cdot p({\rm obs}|\theta),
\end{equation*}
where $p(\theta)$ are the priors of the parameters and $p({\rm obs}|\theta)$ is the likelihood of the entire sample for a given set of parameters $\theta$. 

The likelihood of each set of parameters is computed as follows. 
We write the log-normal probability distribution as: 
\begin{equation*}
p(x|\theta) = \frac{A}{\sqrt{2\pi}\sigma\cdot x} \cdot \exp{[-\frac{(\ln{x}-\mu)^2}{2\sigma^2}]},
\end{equation*}
where $x$ is the Ly$\alpha$ EW or Ly$\alpha$ escape fraction, and $A$ is the normalization parameter: $A=1$ for the Ly$\alpha$ EW distribution, $A=2/[1+{\rm erf}(\frac{-\mu}{\sqrt{2}\sigma})]$ for the Ly$\alpha$ escape fraction distribution \citep{Chen2024}. 
For each galaxy with Ly$\alpha$ emission detections in the sample, the Gaussian measurement uncertainty is: 
\begin{equation*}
p(x)_{{\rm obs,}i} = \frac{1}{\sqrt{2\pi}\sigma_{{\rm obs,}i}} \cdot \exp{[-\frac{(x-x_{{\rm obs,}i})^2}{2\sigma^2_{{\rm obs,}i}}]},
\end{equation*}
where $x_{{\rm obs,}i}$ and $\sigma_{{\rm obs,}i}$ are the observed Ly$\alpha$ EW or Ly$\alpha$ escape fraction and its uncertainty of the $i^{\rm th}$ system. 
Then the individual likelihood of each galaxy with Ly$\alpha$ detection is written as:
\begin{equation*}
p({\rm obs,}i|\theta)_{\rm det} = \int^{\infty}_{0} p(x)_{{\rm obs,}i} \cdot p(x|\theta)\ d\ x.
\end{equation*}
For non-detections of Ly$\alpha$, the absence of OH sky lines in NIRSpec observations significantly improves the completeness of line detection (relative to earlier ground-based observations), the individual likelihood can be simply written as the likelihood of Ly$\alpha$ EW or $f_{{\rm esc,Ly}\alpha}$ below the upper limit:
\begin{equation*}
p({\rm obs,}i|\theta)_{\rm lim} = p(x<x_{3\sigma,i}|\theta),
\end{equation*}
where $x_{3\sigma,i}$ is the $3\sigma$ upper limit of Ly$\alpha$ EW or $f_{{\rm esc,Ly}\alpha}$. 
The total likelihood for each set of parameters is taken as the product of individual likelihoods of all the galaxies in the sample: 
\begin{equation*}
p({\rm obs}|\theta) = \prod_{i} p({\rm obs},i|\theta).
\end{equation*}
Finally, we sample the posteriors of the model parameters using a Markov Chain Monte Carlo (MCMC) approach with the \texttt{emcee} package \citep{Foreman-Mackey2013}. 
For each model parameter, we derive the posterior probability distribution and compute the median value and the marginal $68\%$ credible interval. 

\subsection{The Ly$\alpha$ EW Distribution} \label{sec:lya_ew}

We now derive the Ly$\alpha$ EW distribution and compare with results at $z\sim5-6$ in \citet{Tang2024a}. 
We note that the $z\sim5-6$ Ly$\alpha$ property distributions in \citet{Tang2024a} are derived using spectra obtained with ground-based spectrographs including the Multi Unit Spectroscopic Explorer (MUSE; \citealt{Bacon2010}) at VLT and the DEep Imaging Multi-Object Spectrograph (DEIMOS; \citealt{Faber2003}) at the Keck II telescope. 
Comparison with \textit{JWST}/NIRSpec MSA observations require modest flux conversions due to the different apertures used to extract spectra \citep[e.g.,][]{Bhagwat2024,Chen2024,Nakane2024,Tang2024a}. 
The apertures we used to extract spectra from ground-based facilities have a median diameter of $1.5$~arcsec \citep{Tang2024a}. 
In \citet{Tang2024a} we have derived a conversion factor between the Ly$\alpha$ fluxes measured from MUSE or DEIMOS spectral extractions and NIRSpec MSA shutter spectra (NIRSpec Ly$\alpha$ flux $\simeq0.8\times$ ground-based Ly$\alpha$ flux). 
We multiply the ground-based Ly$\alpha$ fluxes by this factor in order to be consistent with the NIRSpec measurements. 
To quantify the redshift revolution, we divide our sample into three redshift bins: $z=6.5-8.0$ (median $z=7.0$, $153$ galaxies), $z=8.0-10.0$ (median $z=8.8$, $36$ galaxies), and $z=10.0-13.3$ (median $z=11.0$, $12$ galaxies). 
The fitted Ly$\alpha$ EW distribution parameters of each group are presented in Table~\ref{tab:lya_ew_dist_param}. 


\begin{deluxetable}{cccc}
\tabletypesize{\scriptsize}
\tablecaption{Parameters of Ly$\alpha$ EW distributions.}
\tablehead{
Sample & $N_{\rm gal}$ & $e^{\mu}$ (\AA) & $\sigma$ 
}
\startdata
$z=6.5-8.0$ & $153$ & $5.0^{+1.9}_{-1.6}$ & $1.74^{+0.29}_{-0.23}$ \\
$z=8.0-10.0$ & $36$ & $2.7^{+2.0}_{-1.2}$ & $1.59^{+0.50}_{-0.37}$ \\
EGS ($z=6.5-8.0$) & $46$ & $7.1^{+3.9}_{-3.2}$ & $1.98^{+0.52}_{-0.37}$ \\
GOODS + Abell 2744 ($z=6.5-8.0$) & $107$ & $3.9^{+2.1}_{-1.6}$ & $1.66^{+0.39}_{-0.32}$ \\
\enddata
\tabletypesize{\small}
\tablecomments{$N_{\rm gal}$ is the number of galaxies in each subsample. We give the Posterior median and $68\%$ credible interval of the median Ly$\alpha$ EW ($e^{\mu}$) and standard deviation ($\sigma$).}
\label{tab:lya_ew_dist_param}
\end{deluxetable}

\begin{figure}
\includegraphics[width=\linewidth]{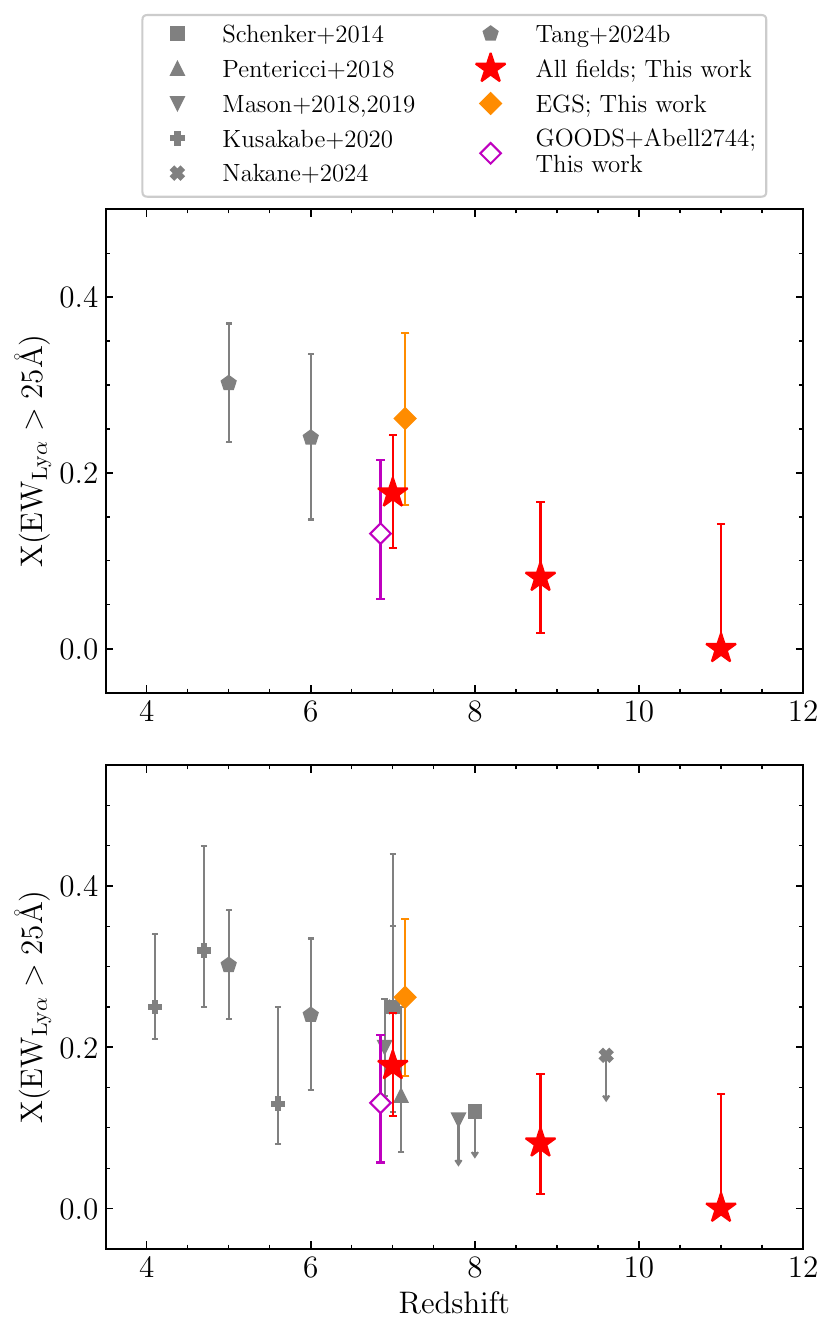}
\caption{Ly$\alpha$ fraction (EW $>25$~\AA) as a function of redshift. We show the NIRSpec samples at $6.5<z<8.0$, $8.0<z<10.0$, and $10.0<z<13.3$ as red stars. We also present the Ly$\alpha$ fractions of galaxies in the EGS (orange filled diamond) and the GOODS + Abell 2744 (open magenta diamond) fields separately, both at $6.5<z<8.0$. The top panel shows the comparison between our Ly$\alpha$ fractions at $z>6.5$ and those at $z\sim5-6$ in \citet[][; grey pentagons]{Tang2024a}. The bottom panel shows the comparison with other literature \citep{Schenker2014,Mason2018,Mason2019a,Pentericci2018,Kusakabe2020,Nakane2024}.}
\label{fig:xlya_ew_z}
\end{figure}

As ground-based investigations focused most on the disappearance of Ly$\alpha$ emitters between $z\simeq6$ and $z\simeq7$, we first investigate whether \textit{JWST} observations suggest a similar picture. 
At $z=6.5-8.0$, we derive the following constraints on the Ly$\alpha$ EW distribution parameters: $\mu=1.60^{+0.32}_{-0.38}$ and $\sigma=1.74^{+0.29}_{-0.23}$. 
This indicates that $18^{+7}_{-6}\%$ of the $z=6.5-8.0$ galaxies show strong Ly$\alpha$ emission with EW $>25$~\AA\ (the so-called Ly$\alpha$ fraction; e.g., \citealt{Stark2010}). 
In the top panel of Figure~\ref{fig:xlya_ew_z}, we compare our Ly$\alpha$ fractions to those derived at $z\sim5$ ($30^{+7}_{-7}\%$) and $z\sim6$ ($24^{+10}_{-9}\%$) in \citet{Tang2024a}. 
The comparison indicates that strong Ly$\alpha$ emission becomes less common from $z\sim5$ to $z\sim7$. 
This is consistent with the trends found in the literature (bottom panel of Figure~\ref{fig:xlya_ew_z}) from ground-based studies \citep[e.g.,][]{Schenker2014,Pentericci2018} and early \textit{JWST} work \citep{Jones2024,Nakane2024}. 


We note that earlier measurements have often focused on systems with $-20.25<{\rm M}_{\rm UV}<-18.75$, narrower than the M$_{\rm UV}$ range of our sample ($-22\lesssim{\rm M}_{\rm UV}\lesssim-15$). 
To evaluate how a different M$_{\rm UV}$ range affects our results, we also derive the Ly$\alpha$ EW distribution and the Ly$\alpha$ fraction of our $z=6.5-8.0$ galaxies with $-20.25<{\rm M}_{\rm UV}<-18.75$ ($79$ galaxies) using the same methodology. 
We find that the Ly$\alpha$ fraction of this subset is $15^{+9}_{-8}\%$, which is consistent with the Ly$\alpha$ fraction of the entire $z=6.5-8.0$ sample ($18^{+7}_{-6}\%$). 
Therefore, we conclude that choosing different M$_{\rm UV}$ ranges (the entire M$_{\rm UV}$ range of our sample or that mostly used in the literature) does not impact our results significantly. 

In spite of the disappearance of Ly$\alpha$ emitters, \textit{JWST} is revealing a small number of galaxies with extremely strong Ly$\alpha$ emission (EW $>100$~\AA) at $z\simeq6.5-8$ (Figure~\ref{fig:lyaew_fesc_z}; see also, e.g., \citealt{Saxena2023,Chen2024,Napolitano2024,Tang2024b,Witstok2024b}). 
Based on our Ly$\alpha$ EW distribution, we infer that $9^{+5}_{-4}\%$ and $4^{+3}_{-2}\%$ of the galaxies at $z=6.5-8.0$ present Ly$\alpha$ EW $>50$~\AA\ and $>100$~\AA, respectively. 
Ly$\alpha$ emission with EW $>100$~\AA\ is sufficiently close to the intrinsic Ly$\alpha$ EWs ($\simeq500$~\AA; e.g., \citealt{Chen2024,Tang2024a}) to suggest a large fraction ($\gtrsim0.2$) of Ly$\alpha$ escapes from these systems.
These galaxies must be located along sightlines where the transmission through the IGM is large ($\gtrsim20\%$ assuming unity transmission through the ISM and CGM). 
It has been shown that this large transmission requires these galaxies are far from the neutral IGM ($\gtrsim1$~pMpc; e.g., \citealt{Mason2020,Saxena2023,Tang2024b,Witstok2024b}). 
As typical bubble sizes decrease at $z\gtrsim8$, we expect these large ionized sightlines (and the extremely strong Ly$\alpha$ emitters they permit) should disappear from our survey volume.

We now consider what current observations reveal about Ly$\alpha$ emission at $z\gtrsim8$. 
Early \textit{JWST} observations took the first steps to constrain the Ly$\alpha$ fraction at $z>8$ based on a handful of spectroscopically-selected galaxies ($N\lesssim10$; e.g., \citealt{Jones2024,Nakane2024,Napolitano2024}). 
The public NIRSpec sample used in this study increases the $z>8$ redshift sample by a factor of $\sim5$ ($N=50$). 
At $z=8.0-10.0$, the Ly$\alpha$ EW distribution parameters are $\mu=0.98^{+0.55}_{-0.58}$ and $\sigma=1.59^{+0.50}_{-0.37}$. 
This indicates a Ly$\alpha$ fraction (EW $>25$~\AA) of $8^{+9}_{-6}\%$, about only half of the Ly$\alpha$ fraction at $z=6.5-8.0$. 
At $z=10.0-13.3$, given the relatively small sample size ($12$ galaxies) we use the statistics for small numbers of events \citep{Gehrels1986} to estimate the Ly$\alpha$ fraction. 
None of the $12$ systems at $z>10$ present Ly$\alpha$ emission with EW $>25$~\AA, resulting in a Ly$\alpha$ fraction of $0^{+14}_{-0}\%$. 
These results suggest that whatever physical process is leading to the disappearance of strong Ly$\alpha$ emitters at $z\simeq7$ continues to $z\simeq9-13$. 
The extremely strong Ly$\alpha$ lines (EW $>100$~\AA) that may provide signposts of large ionized sightlines indeed are extremely rare in existing $z>8$ samples, comprising only $1\%$ of the galaxies at $z>8$. 
We will quantify implications for the evolution in Ly$\alpha$ transmission in Section~\ref{sec:igm_transmission}.

\subsection{The Ly$\alpha$ Escape Fraction Distribution} \label{sec:lya_fesc}

\begin{figure}
\includegraphics[width=\linewidth]{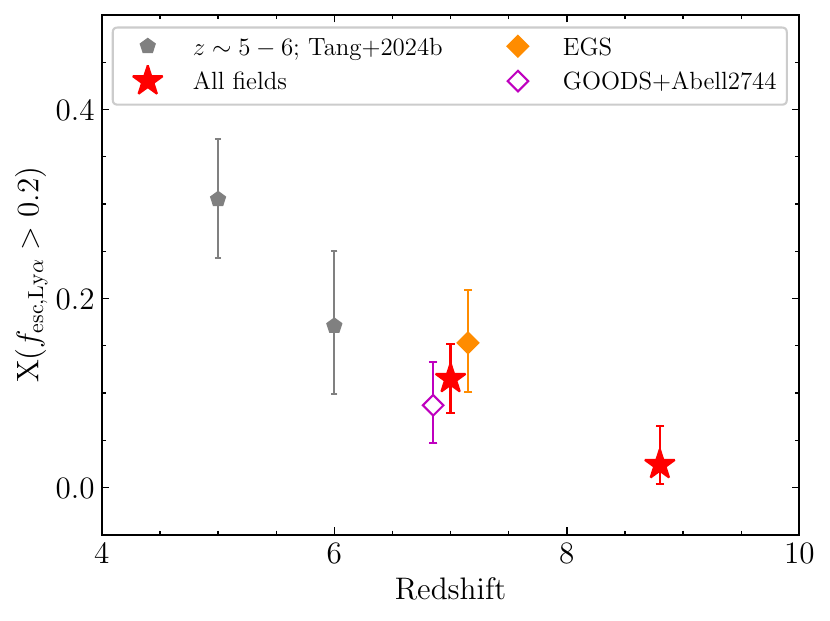}
\caption{The fraction of galaxies with Ly$\alpha$ escape fraction $f_{{\rm esc,Ly}\alpha}>0.2$ as a function of redshift. We show the NIRSpec samples at $6.5<z<8.0$ and $8.0<z<10.0$ as red stars, and show the EGS (orange filled diamond) and the GOODS + Abell 2744 (open magenta diamond) samples at $6.5<z<8.0$ separately. We overplot data at $z\sim5$ and $z\sim6$ from \citep{Tang2024a} as grey pentagons.}
\label{fig:xlya_fesc_z}
\end{figure}

We now consider the evolution in the Ly$\alpha$ escape fraction distribution. 
We consider two redshift bins: $z=6.5-8.0$ ($151$ galaxies) and $z=8.0-10.0$ ($34$ galaxies). 
We do not derive the Ly$\alpha$ escape fraction distribution at $z>10$ because there are only two objects with $f_{{\rm esc,Ly}\alpha}$ constraints. 
The Ly$\alpha$ escape fraction distribution parameters of these two bins are shown in Table~\ref{tab:lya_fesc_dist_param}. 
At $z=6.5-8.0$, the Ly$\alpha$ escape fraction distribution indicates that $12^{+4}_{-4}\%$ of the population shows large Ly$\alpha$ escape fractions ($f_{{\rm esc,Ly}\alpha}>0.2$). 
As a baseline reference, the fractions of galaxies with $f_{{\rm esc,Ly}\alpha}>0.2$ are $31^{+6}_{-6}\%$ at $z\sim5$ and $17^{+7}_{-8}\%$ to $36^{+11}_{-10}\%$ at $z\sim6$ \citep{Chen2024,Tang2024a}. 
The results suggest that the fraction of galaxies with large Ly$\alpha$ escape fractions decreases by a factor of $2-3$ from $z\sim5$ to $z\sim7$ (Figure~\ref{fig:xlya_fesc_z}). 
The fraction of galaxies with large Ly$\alpha$ transmission decreases further at $z=8.0-10.0$, with only $2^{+4}_{-2}\%$ of the population presenting $f_{{\rm esc,Ly}\alpha}>0.2$ (Figure~\ref{fig:xlya_fesc_z}). 
We will discuss implications of the evolving Ly$\alpha$ transmission in Section~\ref{sec:igm_transmission}. 


\begin{deluxetable}{cccc}
\tabletypesize{\scriptsize}
\tablecaption{Parameters of Ly$\alpha$ escape fraction distributions.}
\tablehead{
Sample & $N_{\rm gal}$ & $e^{\mu}$ & $\sigma$ 
}
\startdata
$z=6.5-8.0$ & $151$ & $0.04^{+0.01}_{-0.01}$ & $1.29^{+0.16}_{-0.15}$ \\
$z=8.0-10.0$ & $34$ & $0.03^{+0.01}_{-0.01}$ & $0.93^{+0.19}_{-0.16}$ \\
EGS ($z=6.5-8.0$) & $46$ & $0.06^{+0.02}_{-0.01}$ & $1.23^{+0.19}_{-0.16}$ \\
GOODS + Abell 2744 ($z=6.5-8.0$) & $105$ & $0.04^{+0.01}_{-0.01}$ & $1.13^{+0.17}_{-0.15}$ \\
\enddata
\tabletypesize{\small}
\tablecomments{$N_{\rm gal}$ is the number of galaxies in each subsample. We give the Posterior median and $68\%$ credible interval of the median Ly$\alpha$ escape fraction ($e^{\mu}$) and standard deviation ($\sigma$).}
\label{tab:lya_fesc_dist_param}
\end{deluxetable}


\subsection{Field-to-Field Variations} \label{sec:lya_variation}

In the final portion of this section, we explore the field-to-field variations of the distributions of Ly$\alpha$ emitters at $z>6.5$. 
If there are large ($\gtrsim1$~pMpc) ionized sightlines corresponding to early intergalactic bubbles, the counts of Ly$\alpha$ emitters will vary greatly in different fields.

\begin{figure*}
\includegraphics[width=\linewidth]{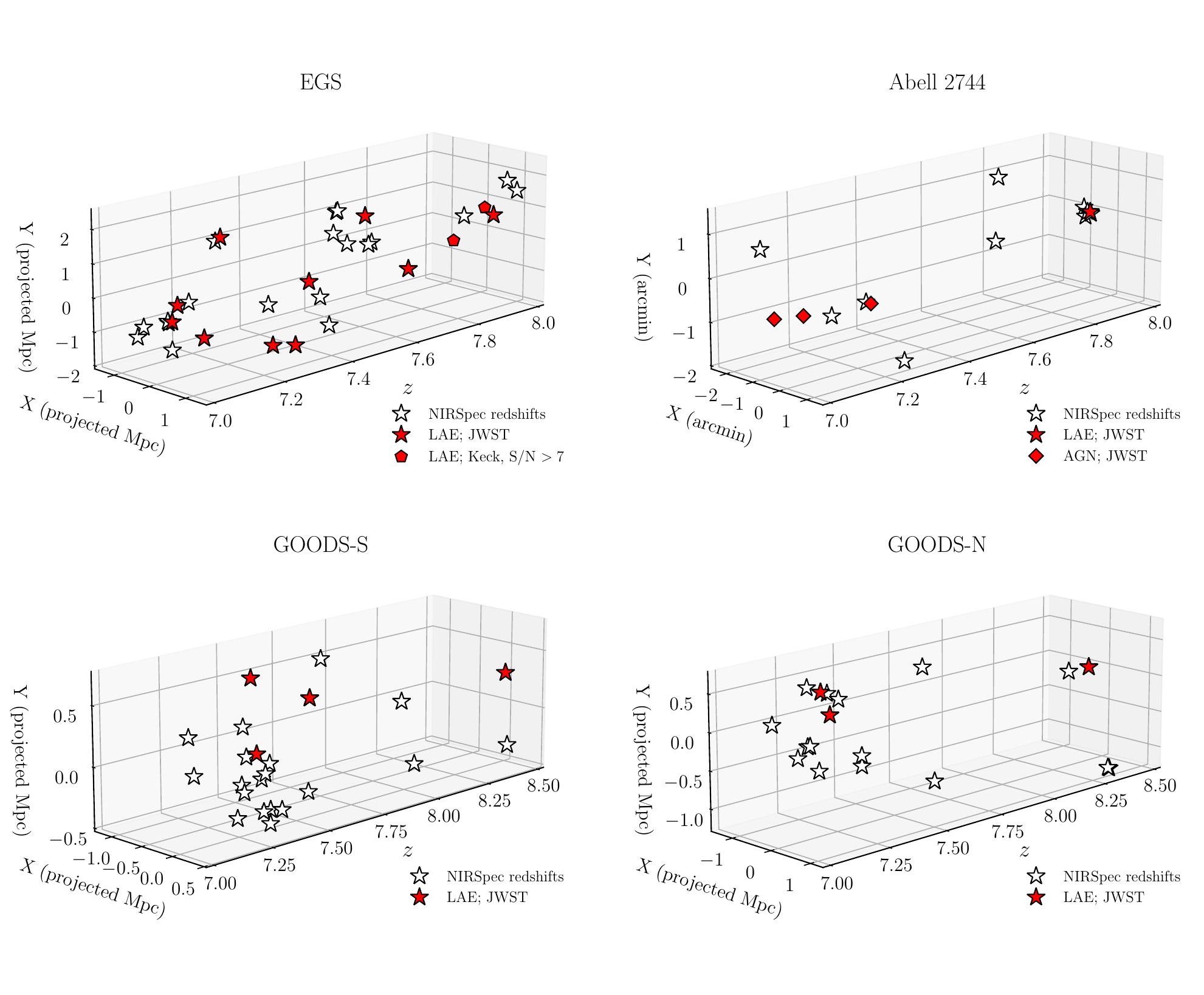}
\caption{Spatial distribution of the spectroscopically confirmed galaxies (stars) at $z\simeq7-8$ identified from public NIRSpec datasets in the EGS (top left), Abell 2744 (top right), GOODS-South (bottom left), and GOODS-North (bottom right) fields. We mark Ly$\alpha$ emitters as red filled stars. We also show AGNs as diamonds. In the EGS field, we overplot high S/N ($>7$) Ly$\alpha$ detections from ground-based observations \citep{Oesch2015,Tilvi2020,Jung2022} as red pentagons.}
\label{fig:space_dist}
\end{figure*}

We show the spatial distributions of spectroscopically-confirmed galaxies at $z\simeq7-8$ in the four fields sampled by \textit{JWST} observations (EGS, Abell 2744, GOODS-S, and GOODS-N) in Figure~\ref{fig:space_dist}. 
We also include the small number of Ly$\alpha$ emitters with $>7\sigma$ Ly$\alpha$ detections identified from ground-based telescope observations that have yet to be observed with \textit{JWST} \citep{Oesch2015,Tilvi2020,Jung2022}. 
We see clearly that there is significant variance in the counts of Ly$\alpha$ emitters in the four observed fields. 
The EGS now has $12$ Ly$\alpha$ detections in the redshift range $7<z<8$, far greater than what has been reported in the other three fields. 
In spite of considerable observational investment, the GOODS-S and GOODS-N fields have only revealed $4$ and $3$ Ly$\alpha$ emitting galaxies at $z\simeq7-8$, respectively. 
The Abell 2744 field currently has only $4$ Ly$\alpha$ detections in the same redshift range, and $3$ of them appear to be broad line AGNs \citep{Furtak2023a,Kocevski2024}. 


To quantify whether the excess of Ly$\alpha$ detections in EGS is statistically meaningful with existing data, we calculate the distribution of Ly$\alpha$ properties in each field. 
The EW and escape fraction distribution parameters are presented in Table~\ref{tab:lya_ew_dist_param} and Table~\ref{tab:lya_fesc_dist_param}, respectively. 
In the EGS, from our inferred Ly$\alpha$ EW distribution we find that $26^{+10}_{-10}\%$ of the $z=6.5-8.0$ galaxies in this field present strong (EW $>25$~\AA) Ly$\alpha$ emission. 
This is two times greater than the Ly$\alpha$ fraction inferred in the other three fields (Figure~\ref{fig:xlya_ew_z}). 
The Ly$\alpha$ escape fraction distribution shows a similar result, with the EGS distribution displaying a larger fraction of high escape fraction ($f_{{\rm esc,Ly}\alpha}>0.2$) galaxies (Figure~\ref{fig:xlya_fesc_z}). 

Also of note in Figure~\ref{fig:xlya_ew_z} and Figure~\ref{fig:xlya_fesc_z} is the comparison of the field-dependent Ly$\alpha$ fraction at $z\simeq7$ and $z\simeq6$. 
The EGS shows no evidence for a significant downturn in Ly$\alpha$ emission that is seen in the three other fields targeted to date. 
While uncertainties are still significant, current results suggest that the impact of the damping wing from the neutral IGM may be significantly reduced in the EGS field at $z\simeq7-8$, as expected in regions with large ionized sightlines. 
Of the four fields targeted so far with \textit{JWST}, it does appear as if the EGS may be host to the largest ionized regions at $z\simeq7-8$. 
Statistical uncertainties will be greatly reduced as larger rest-frame UV spectroscopic samples are obtained.


\section{L\lowercase{y}$\alpha$ Emission Transmission at $\lowercase{z}>6.5$ and Implications for the Early IGM} \label{sec:igm_transmission}

In Section~\ref{sec:lya_evolution}, we quantified the disappearance of galaxies with large Ly$\alpha$ EWs and large escape fraction of Ly$\alpha$ from $z\simeq5$ to $z\gtrsim10$. 
In this section, we will constrain the evolution in the transmission of Ly$\alpha$ photons that is required to explain the observations (Section~\ref{sec:lya_transmission}). 
Assuming this is driven by damping wing attenuation from the neutral IGM, we will estimate the neutral hydrogen fraction in the IGM ($x_{\rm HI}$) at $z>6.5$ following similar methods to that used in the literature \citep[e.g.,][]{Mason2018,Mason2019a} (Section~\ref{sec:neutral_fraction}). 
Here our focus will primarily be on the very early stages of reionization ($z\gtrsim8.5$) that \textit{JWST} observations are now probing.

\subsection{The Evolving Transmission of Ly$\alpha$ Emission} \label{sec:lya_transmission}

To derive the evolution in the transmission of Ly$\alpha$, we compare the rest-frame Ly$\alpha$ EWs at $z>6.5$ with the Ly$\alpha$ EW distribution at $z\sim5$. 
We will assume that changes in Ly$\alpha$ are driven by evolution in the IGM transmission ($\mathcal{T}_{\rm IGM}$), but we will discuss the likelihood of additional contributions from evolving galaxy properties below. 
Recent studies have suggested an endpoint of reionization at $z\sim5.3$ \citep[e.g.,][]{Becker2021,Bosman2022,Spina2024,Zhu2024}. 
It is thus standard to use the Ly$\alpha$ EW distribution at $z\simeq5$ as the baseline against which higher redshift samples are assessed. 
We will make use of the $z\simeq5$ Ly$\alpha$ EW distribution that was recently derived in \citet{Tang2024a} using ground-based spectroscopy together with \textit{JWST}/NIRCam photometry. 
As discussed in \citet{Tang2024a} and in Section~\ref{sec:lya_evolution}, comparison of the $z\simeq5-6$ distributions with $z\gtrsim6.5$ \textit{JWST}/NIRSpec observations requires correction for the aperture mismatch between the ground-based spectra and the NIRSpec microshutter. 
For the $z\simeq5-6$ ground-based observations used in this study, this aperture correction is on average given by $F_{{\rm Ly}\alpha,{\rm NIRSpec}}\simeq0.8\times F_{{\rm Ly}\alpha,{\rm ground}}$ (see \citealt{Tang2024a}). 
After applying this correction, the $z\simeq5$ Ly$\alpha$ EW distribution, $p_{z\sim5}({\rm EW})$, can be modeled by a log-normal distribution with parameters $\mu=2.38^{+0.28}_{-0.31}$ and $\sigma=1.64^{+0.23}_{-0.19}$. 
Since we are interested in the transmission relative to $z\simeq5$, we assume the average IGM transmission of Ly$\alpha$ at $z\sim5$ is unity ($\mathcal{T}_{{\rm IGM},z\sim5}=1$) and then write a forward model of Ly$\alpha$ EW distribution as $p({\rm EW}|\mathcal{T}_{\rm IGM})=p_{z\sim5}(\frac{{\rm EW}}{\mathcal{T}_{\rm IGM}})$. 
We note that the residual H~{\small I} in the ionized IGM at $z\simeq5$ does impact Ly$\alpha$ transmission (see \citealt{Tang2024a}), but this does not affect our results since we are only interested in computing the \textit{relative} transmission as the IGM becomes increasingly neutral at $z\gtrsim5.3$.

Equipped with the forward model of Ly$\alpha$ EW distribution, we infer the average IGM transmission of Ly$\alpha$ at $z\gtrsim6.5$ (relative to $z\sim5$) using a Bayesian approach \citep[e.g.,][]{Mason2018}. 
Based on Bayes' Theorem, we derive the posterior probability distribution of $\mathcal{T}_{\rm IGM}$ inferred from Ly$\alpha$ EWs as: 
\begin{equation*}
p(\mathcal{T}_{\rm IGM}|{\rm EW}) \propto p(\mathcal{T}_{\rm IGM})\cdot \prod_{i} p({\rm EW}_i|\mathcal{T}_{\rm IGM}). 
\end{equation*}
Here $p(\mathcal{T}_{\rm IGM})$ is the prior on the IGM transmission, and we assume a uniform prior with $0<\mathcal{T}_{\rm IGM}<1$. 
The likelihood of the entire NIRSpec sample for a given $\mathcal{T}_{\rm IGM}$ is written as $\prod_{i} p({\rm EW}_i|\mathcal{T}_{\rm IGM})$, which is taken as the product of the individual likelihood of all the galaxies in the sample. 
For each galaxy with a Ly$\alpha$ detection, the likelihood function is given by:
\begin{equation*}
p({\rm EW}_i|\mathcal{T}_{\rm IGM})_{\rm det} = \int^{\infty}_{0} d{\rm EW}\ \frac{e^{-\frac{({\rm EW}-{\rm EW}_i)^2}{2\sigma^2_i}}}{\sqrt{2\pi}\sigma_i}\ p({\rm EW}|\mathcal{T}_{\rm IGM}), 
\end{equation*}
where EW$_i$ and $\sigma_i$ are the measured Ly$\alpha$ EW and the $1\sigma$ uncertainty of the $i_{\rm th}$ galaxy. 
For each source where Ly$\alpha$ is undetected, the individual likelihood is as follows: 
\begin{eqnarray*}
p({\rm EW}_i|\mathcal{T}_{\rm IGM})_{\rm lim} &=& \int^{\infty}_{0} d{\rm EW}\ \frac{1}{2} {\rm erfc}(\frac{{\rm EW}-{\rm EW}_{3\sigma,i}}{\sqrt{2}\sigma_i})\cdot \\
&& p({\rm EW}|\mathcal{T}_{\rm IGM}), 
\end{eqnarray*}
where ${\rm EW}_{3\sigma,i}$ is the $3\sigma$ upper limit of Ly$\alpha$ EW, and erfc is the complementary error function. 
We then sample the posterior of the IGM transmission using a MCMC approach with the \texttt{emcee} package. 
We derive the posterior probability distribution of $\mathcal{T}_{\rm IGM}$ and compute the median value and the marginal $68\%$ credible interval. 

We now derive the average transmission of Ly$\alpha$ (relative to $z\simeq5$) for galaxies in three redshift ranges: $z=6.5-8.0$ (median $z=7.0$), $z=8.0-10.0$ (median $z=8.8$), and $z=10.0-13.3$ (median $z=11.0$). 
At $z=6.5-8.0$, we require an IGM transmission $\mathcal{T}_{\rm IGM}=0.51^{+0.11}_{-0.09}$ to reproduce our observations. 
This suggests that the average IGM transmission of Ly$\alpha$ at $z=6.5-8.0$ is about half that at $z\sim5$. 
The average IGM transmission decreases at yet higher redshifts, with $\mathcal{T}_{\rm IGM}=0.26^{+0.15}_{-0.10}$ at $z=8.0-10.0$ and $\mathcal{T}_{\rm IGM}=0.16^{+0.19}_{-0.09}$ at $z=10.0-13.3$. 
These findings indicate that the transmission of Ly$\alpha$ photons decreases by $4\times$ between $z\simeq5$ and $z\simeq9$. 
This result is now clearly established with \textit{JWST} spectroscopy (see also \citealt{Tang2023,Chen2024}), marking a significant step beyond the uncertainties that were inherent in ground-based infrared campaigns (as described in Section~\ref{sec:introduction}). 

Observational work has long sought to determine how much of the evolution in Ly$\alpha$ is driven by changes in galaxies and how much is due to the IGM. 
As our understanding of the demographics of the galaxy population has improved, it has become clear that galaxies are unlikely to be the primary factor driving this evolution. 
Indeed at $z\gtrsim3$, the physical properties of galaxies shift towards higher sSFR and lower dust attenuation, implying a \textit{higher} intrinsic production efficiency and escape of Ly$\alpha$ photons from galaxies \citep[e.g.,][]{Stark2010,Stefanon2022,Topping2022,Topping2024b,Cullen2023,Morales2024}. 
Measurements of low ionization absorption lines at $z\simeq4-5$ suggest that the covering fraction of neutral gas in the ISM and CGM of galaxies decreases toward earlier times \citep[e.g.,][]{Jones2012,Du2018,Pahl2020}, likely allowing a larger fraction of Ly$\alpha$ radiation to be transmitted through galaxies. 
Collectively these observations indicate that galaxies are evolving in a manner that allows more Ly$\alpha$ to escape at earlier epochs, consistent with the evolution in Ly$\alpha$ EW distribution at $z\gtrsim3$ \citep[e.g.,][]{Stark2011,Cassata2015,ArrabalHaro2018,deLaVieuville2020,Kusakabe2020,Tang2024a}

\textit{JWST} continuum spectroscopy is now allowing new avenues of probing the influence of galaxies on the transfer of Ly$\alpha$ photons, building on the work described above. 
Recent attention has focused on the discovery of strong absorption around the Ly$\alpha$ break in $z>5$ galaxies, likely by dense neutral gas in the ISM and CGM \citep[e.g.,][]{Cameron2024,Chen2024,Hainline2024,Heintz2024a,Heintz2024c}. 
Of course the presence of such damped Ly$\alpha$ spectra are not surprising, as they are seen in a large fraction of $z\simeq3$ galaxy spectra \citep[e.g.,][]{Shapley2003}. 
Based on existing \textit{JWST} spectroscopy, there is no indication that gas column densities required to produce the damped Ly$\alpha$ profiles are significantly different than at lower redshifts \citep[e.g.][]{Reddy2016}, or evolve strongly with redshift $z\sim5-11$ (Mason et al. in prep.). 
The absence of strong evolution toward large H~{\small I} column densities in the CGM of early galaxies is consistent with indications for lower neutral gas covering fractions at $z\gtrsim3$ described above. 
While there have long been suggestions from quasar absorption line studies that the CGM may become more neutral at $z\gtrsim5$ \citep{Becker2019,Cooper2019,Christensen2023,Sebastian2024}, it is thought that this effect is not able to dominate the evolving Ly$\alpha$ transmission seen in galaxy spectra. 
An increasingly neutral IGM is the most likely driver \citep{Mesinger2015}. 
As higher resolution \textit{JWST} continuum spectra are obtained, it will be possible to revisit the quantitative impact of self-shielding absorption systems on Ly$\alpha$ emission. 
In what follows, we will consider what neutral hydrogen fraction is required for the damping wing attenuation to reproduce the Ly$\alpha$ emission spectra. 

\subsection{Implications for the Early IGM Neutral Fraction} \label{sec:neutral_fraction}

We now quantify the neutral hydrogen fraction required to explain the decrease in the Ly$\alpha$ transmission between $z\sim5$ to $z>6.5$, making the standard assumption that an increasingly neutral IGM is the primary source of opacity to Ly$\alpha$ photons.
In addition to the attenuation provided by the damping wing of the neutral IGM \citep{Miralda-Escude1998}, it is standard to also account for the effect of infalling intergalactic gas onto galaxies. 
Even if the regions surrounding $z\gtrsim5.3$ galaxies have been reionized, the IGM density is high enough that the residual neutral hydrogen resonantly scatters photons blueward of the Ly$\alpha$ resonance \citep{Gunn1965}. 
If the gas is infalling, Ly$\alpha$ will also be scattered on the red side of the systemic redshift \citep[e.g.,][]{Santos2004,Dijkstra2007,Laursen2011,Mason2018}. 

The optical depth provided by the IGM to the Ly$\alpha$ profile is computed by combining these two effects as a function of velocity ($v$) with respect to line center:
\begin{equation*}
\tau_{\rm IGM}(v) = \tau_{\rm HII}(v) + \tau_{\rm DW}(v),
\end{equation*}
where $\tau_{\rm HII}$ is the optical depth due to scattering from residual neutral hydrogen in the ionized bubble around the galaxy, and $\tau_{\rm DW}$ is the damping wing optical depth from the neutral IGM.
To evaluate $\tau_{\rm HII}$, we follow the approach described in \citet{Mason2018}. 
In this model, the IGM is assumed to be infalling at the circular velocity ($v_c$) of the halo, and the halo mass ($M_h$) is estimated from the redshift and M$_{\rm UV}$ of the galaxy using the M$_{\rm UV}-M_h$ relations in \citet{Mason2015}. 
The reader is directed to that paper for details, but in this approach, more luminous galaxies will have attenuation from infalling gas extending to larger velocities on the red side of line center. 
For simplicity, we assume $\tau_{\rm HII}=\infty$ at $v\le v_c$ and $\tau_{\rm HII}=0$ at $v>v_c$. 
To give a quantitative example, this infall model will cause a galaxy with M$_{\rm UV}=-19$ to have zero transmission of its Ly$\alpha$ at velocities below $140$~km~s$^{-1}$. 
We will discuss the impact of infall on the evolution of Ly$\alpha$ emitters at $z\gtrsim6.5$ later in the section.

The IGM damping wing attenuates the Ly$\alpha$ profile differently as a function of wavelength.
To compute the damping wing optical depth ($\tau_{\rm DW}$) provided to Ly$\alpha$ photons at observed wavelength $\lambda_{\rm obs}$, we follow the prescription described in \citet{Mason2020}: 
\begin{equation*}
\tau_{\rm DW}(\lambda_{\rm obs}) = \int^{z_s}_{z_{\rm reion}} dz\ c \frac{dt}{dz}\ x_{\rm HI}(z) n_{\rm H}(z) \sigma_{\alpha}(\frac{\lambda_{\rm obs}}{1+z},T),
\end{equation*}
integrating the attenuation provided to a Ly$\alpha$ photon emitted at wavelength $\lambda_{\rm em}$ from a galaxy at redshift $z_s$. 
Here the emitted wavelength is the rest-frame equivalent of the observed wavelength in equation: $\lambda_{\rm em}=\lambda_{\rm obs}/(1+z_s)$.
The integral accounts for the Ly$\alpha$ opacity as the line emission is redshifted through the IGM. The number density of scattering H~{\small I} particles in each differential redshift bin is given by the product of the IGM neutral fraction and the number density of hydrogen atoms, $x_{\rm HI}(z)$ and $n_{\rm H}(z)$, respectively. 
The Ly$\alpha$ scattering cross section is given by $\sigma_\alpha$, defined in the rest-frame of the scattering particles as a function of wavelength and H~{\small I} kinetic temperature (see \citealt{Dijkstra2014} for the complete functional form). 
As Ly$\alpha$ photons are redshifted, they move to larger wavelengths where the cross section is greatly reduced. 
In theory, the integral is computed from the redshift where the IGM becomes somewhat neutral ($z_{\rm reion}\simeq5.3$) to the redshift of the galaxy. 
In practice, the damping wing optical depth depends negligibly on the value chosen for $z_{\rm reion}$ given that most of the opacity is from H~{\small I} at redshifts closer to the galaxy redshift ($z_s\gtrsim6.5$ in our sample) where the scattering cross section is still significant.

We consider a galaxy situated at the center of an ionized bubble, where the distance to the neutral IGM is given by $R_b$ (i.e., the radius of the ionized bubble). 
For the ionized bubble sizes, we adopt the size distribution $p(R_b|z_s,{\rm M}_{\rm UV},x_{\rm HI})$ in \citet{Lu2024} given the source redshift $z_s$, M$_{\rm UV}$, and the IGM neutral fraction $x_{\rm HI}$. 
The ionized bubble size distributions in \citet{Lu2024} are derived from the semi-numerical cosmological simulation code \texttt{21cmfast} \citep{Mesinger2011,Sobacchi2014,Mesinger2016}. 
We will adopt the source model where reionization is driven by numerous UV-faint galaxies (the ``gradual'' model), but our main conclusions are not very sensitive to this choice.
We compute the damping wing optical depth in two steps, following approach in \citealt{Mason2020}.
We first integrate within the ionized bubble from $z_s$ to $z_{\rm ion}$, where $z_{\rm ion}$ is the boundary of the ionized bubble at distance $R_b$ from the source galaxy and hydrogen is assumed to be ionized within the bubble. 
We next integrate outside the ionized bubble from $z_{\rm ion}$ to $z_{\rm reion}$. 
Because the damping wing optical depth is dominated by the distance to the first neutral patch of hydrogen \citep[e.g.,][]{Mesinger2008}, we assume the IGM is neutral ($x_{\rm HI}=1$) outside the bubble. 

Given the total optical depth to Ly$\alpha$, we model the IGM transmission of the Ly$\alpha$ emission emerging from a galaxy with M$_{\rm UV}$ at redshift $z_s$ as \citep[e.g.,][]{Dijkstra2011,Mason2018}: 
\begin{eqnarray*}
\mathcal{T}_{\rm IGM}&&(z_s,{\rm M}_{\rm UV},x_{\rm HI}) = \int^{\infty}_{0} dR_b\ p(R_b|z_s,{\rm M}_{\rm UV},x_{\rm HI})\cdot \\ 
&& \int^{\infty}_{-\infty} dv\ J_{\alpha}(v)\cdot\exp{[-\tau_{\rm IGM}(z_s,{\rm M}_{\rm UV},x_{\rm HI},R_b,v)]}. 
\end{eqnarray*}
$J_{\alpha}(v)$ is the normalized Ly$\alpha$ velocity profile emerging from the ISM and CGM of the host galaxy prior to interaction with the neutral IGM outside the ionized bubble.
The term $\int^{\infty}_{-\infty} dv\ J_{\alpha}(v)\cdot\exp{[-\tau_{\rm IGM}(z_s,{\rm M}_{\rm UV},x_{\rm HI},R_b,v)]}$ computes the IGM transmission of the Ly$\alpha$ velocity profile for a galaxy in an ionized bubble with radius $R_b$, accounting for both infall and the damping wing.
Then we integrate this transmission with the size distribution $p(R_b)$ given the source redshift and M$_{\rm UV}$ to calculate the average IGM transmission for a given $x_{\rm HI}$. 
Here we adopt the composite Ly$\alpha$ velocity profile of Ly$\alpha$ emitting galaxies at $z\simeq5$ in \citet{Tang2024a} as $J_{\alpha}(v)$. 
We refer readers to \citet{Tang2024a} for details of the $z\simeq5$ Ly$\alpha$ emitter sample. 
The composite has a peak velocity offset $\Delta v_{{\rm Ly}\alpha}\simeq230$~km~s$^{-1}$, and a FWHM $\simeq300$~km~s$^{-1}$. Given the M$_{\rm{UV}}$ range of our sample and the large velocity offsets at $z\simeq5$, the effects of infall are not predicted to contribute to the declining IGM transmission through reionization era. 

We apply the IGM transmission models to the Ly$\alpha$ EW distributions at $z\sim5$ \citep{Tang2024a} to derive the forward model of Ly$\alpha$ EW distribution $p({\rm EW}|x_{\rm HI})=p_{z\sim5}(\frac{{\rm EW}}{\mathcal{T}_{\rm IGM}})$. 
We then infer the IGM neutral fraction at $z>6.5$ using a similar Bayesian approach to that described in \citet{Mason2018}. 
The posterior probability distribution of $x_{\rm HI}$ is derived from NIRSpec Ly$\alpha$ EW measurements: 
\begin{equation*}
p(x_{\rm HI}|{\rm EW}) \propto p(x_{\rm HI})\cdot \prod_{i} p({\rm EW}_i|x_{\rm HI}), 
\end{equation*}
where $p(x_{\rm HI})$ is the prior of IGM neutral fraction (assuming a uniform prior $0\le x_{\rm HI}\le1$), and $\prod_{i} p({\rm EW}_i|x_{\rm HI})$ is the likelihood of the entire NIRSpec sample for a given $x_{\rm HI}$. 
The likelihood function of each galaxy is computed as the same way as when deriving the average IGM transmission (see Section~\ref{sec:lya_transmission}). 
Finally, we sample the posterior of the IGM neutral fraction using a MCMC approach with the \texttt{emcee} package. 
We compute the median value and the marginal $68\%$ credible interval of $x_{\rm HI}$ from the posterior probability distribution. 


\begin{figure*}
\centering
\includegraphics[width=0.75\linewidth]{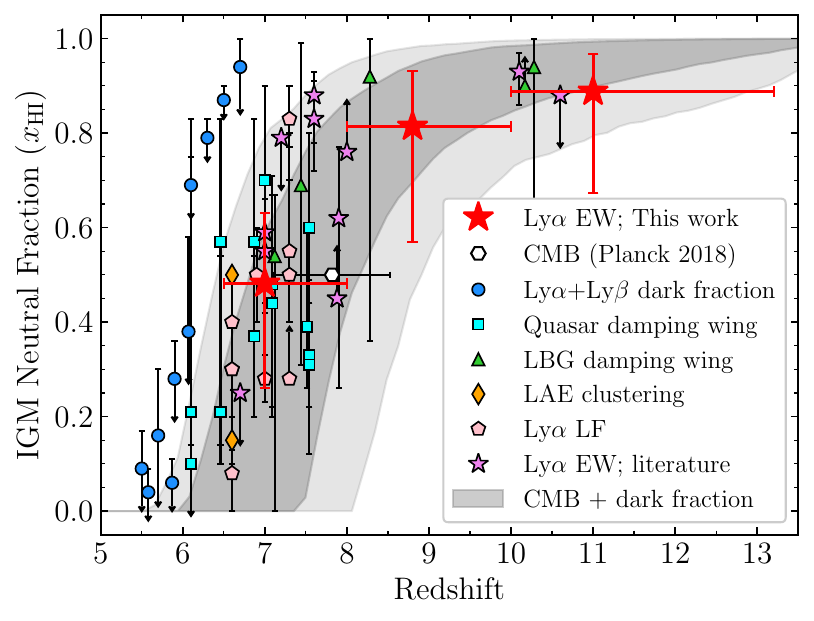}
\caption{Evolution of IGM neutral fraction ($x_{\rm HI}$) in the reionization era. The IGM neutral fractions derived from our $z>6.5$ NIRSpec sample are shown by red stars. We show the constraints derived from CMB optical depth and Ly$\alpha$ and Ly$\beta$ forest dark pixel fraction \citep{Mason2019b} as grey shaded regions (dark grey: $68\%$ percentiles; light grey: $95\%$ percentiles). We overplot the IGM neutral fractions derived from multiple observational probes in literature: CMB optical depth (white hexagon; \citealt{Planck2020}), Ly$\alpha$ and Ly$\beta$ forest dark pixel fraction (blue circles; \citealt{McGreer2015,Jin2023}), Ly$\alpha$ damping wing absorption of quasars (cyan squares; \citealt{Banados2018,Davies2018,Wang2020,Yang2020,Greig2022,Durovcikova2024}), Ly$\alpha$ damping wing absorption of Lyman break galaxies (green triangles; \citealt{Curtis-Lake2023,Hsiao2023,Umeda2024}), the clustering of LAEs (orange diamond; \citealt{Sobacchi2015,Ouchi2018}), Ly$\alpha$ luminosity function (pink pentagon; \citealt{Ouchi2010,Konno2014,Konno2018,Zheng2017,Inoue2018,Goto2021,Morales2021,Ning2022}), and Ly$\alpha$ EW (violet stars; \citealt{Mason2018,Mason2019a,Hoag2019,Whitler2020,Bolan2022,Bruton2023,Morishita2023,Nakane2024}).}
\label{fig:xhi}
\end{figure*}

In Figure~\ref{fig:xhi}, we show the inferred IGM neutral hydrogen fractions as a function of redshift (red stars). 
At $z=6.5-8.0$, we derive an IGM neutral fraction of $x_{\rm HI}=0.48^{+0.15}_{-0.22}$. 
This is broadly consistent with inferences from prior to \textit{JWST}, but the new results are now based on spectroscopically-confirmed samples with much-improved reliability in the flux measurements. 
One of the primary advantages of \textit{JWST} is the ability to extend measurements to $z\gtrsim8$. 
We find that the downturn in Ly$\alpha$ transmission suggests very large neutral fractions at $z=8.0-10.0$ ($x_{\rm HI}=0.81^{+0.12}_{-0.24}$) and $z=10.0-13.3$ ($x_{\rm HI}=0.89^{+0.08}_{-0.21}$). 
These values are derived from the largest $z\gtrsim8$ samples to date, but are consistent with recent measurements in the literature \citep[e.g.,][]{Bruton2023,Curtis-Lake2023,Hsiao2023,Nakane2024,Umeda2024}

The extremely large neutral fraction that \textit{JWST} is revealing at $z\simeq8-13$ may seem surprising in light of the large density of UV photons that appears to be in place at $z\gtrsim10$ \citep[e.g.,][]{Adams2023,Bouwens2023,Harikane2023,Donnan2024,Finkelstein2024}. 
It has been shown that these new \textit{JWST} measurements may indicate that reionization had an earlier start than we previously expected, with a non-negligible fraction of the IGM ionized at $z\gtrsim10$. 
\citet{Gelli2024} demonstrate that in some extreme models, the neutral fractions may already be $x_{\rm{HI}}=0.7$ at $z\simeq9$. 
The neutral fractions depend on source assumptions, with the lowest neutral fractions expected in the \citet{Gelli2024} models where the Lyman continuum escape fraction is exponentially boosted during strong bursts of star formation. 
If these models are correct, we should find strong Ly$\alpha$ emitters (EW $>100$~\AA) at $z\gtrsim9$ as spectroscopic samples increase in size. 
At present, current results hint at a significant reduction in Ly$\alpha$ transmission at $z\gtrsim9$, which is more consistent with the less extreme ionizing photon assumptions in \citet{Gelli2024}. 
While uncertainties remain sizable, the sample statistics will be greatly improved in the future, allowing \textit{JWST} Ly$\alpha$ studies to provide one of our only windows on the earliest phases of reionization.


\section{The Galaxy Environment associated with strong $\lowercase{z}\gtrsim7$ L\lowercase{y}$\alpha$ emitters} \label{sec:lya_association}

In the previous sections, we have demonstrated that \textit{JWST} is capable of identifying strong Ly$\alpha$ emitters at redshifts ($z\gtrsim7$) where the IGM is significantly neutral. 
We now consider whether these systems tend to probe overdense regions that may be capable of carving large ionized bubbles. 
We focus on Ly$\alpha$ detections in the footprint of the two GOODS fields that have been targeted with the NIRCam grism by the First Reionization Epoch Spectroscopically Complete Observations (FRESCO; \citealt{Oesch2023}) team. 
Given the blind nature of grism follow-up, these areas will offer the highest level of completeness (at least for galaxies that have strong [O~{\small III}] emission). 
FRESCO observed an area of $62$~arcmin$^2$ in each of the GOODS fields using the F444W slitless grism, providing wavelength coverage that is ideal for confirming galaxies at $6.8<z<9.0$ via strong [O~{\small III}]~$\lambda4959$ and $5007$ emission. 
Here we make use of the [O~{\small III}] emitter catalog released by the FRESCO team in \citet{Meyer2024}. 

In the area covered by FRESCO, there are currently four known Ly$\alpha$ emitters with large EWs ($>50$~\AA) and large Ly$\alpha$ transmissions ($f_{{\rm esc,Ly}\alpha}>0.2$) at $z\simeq7.0-8.5$ (JADES-1129, $z=7.087$; JADES-13041, $z=7.090$; JADES-13682, $z=7.275$; JADES-1899, $z=8.279$; Table~\ref{tab:sources}). 
Their Ly$\alpha$ escape fractions and EWs are among the top $10\%$ of the $f_{{\rm esc,Ly}\alpha}$ and Ly$\alpha$ EW distributions in the JADES fields at $z\simeq7-8$ (Section~\ref{sec:lya_evolution}). 
We note that JADES-1129 and JADES-13041 are both at roughly the same redshift ($z\simeq7.1)$ in GOODS-N.
Our goal is to investigate whether strong spectroscopic overdensities are required to facilitate sightlines with such strong Ly$\alpha$ emission (see also \citealt{Witstok2024a,Witstok2024b}).


\begin{figure*}
\includegraphics[width=\linewidth]{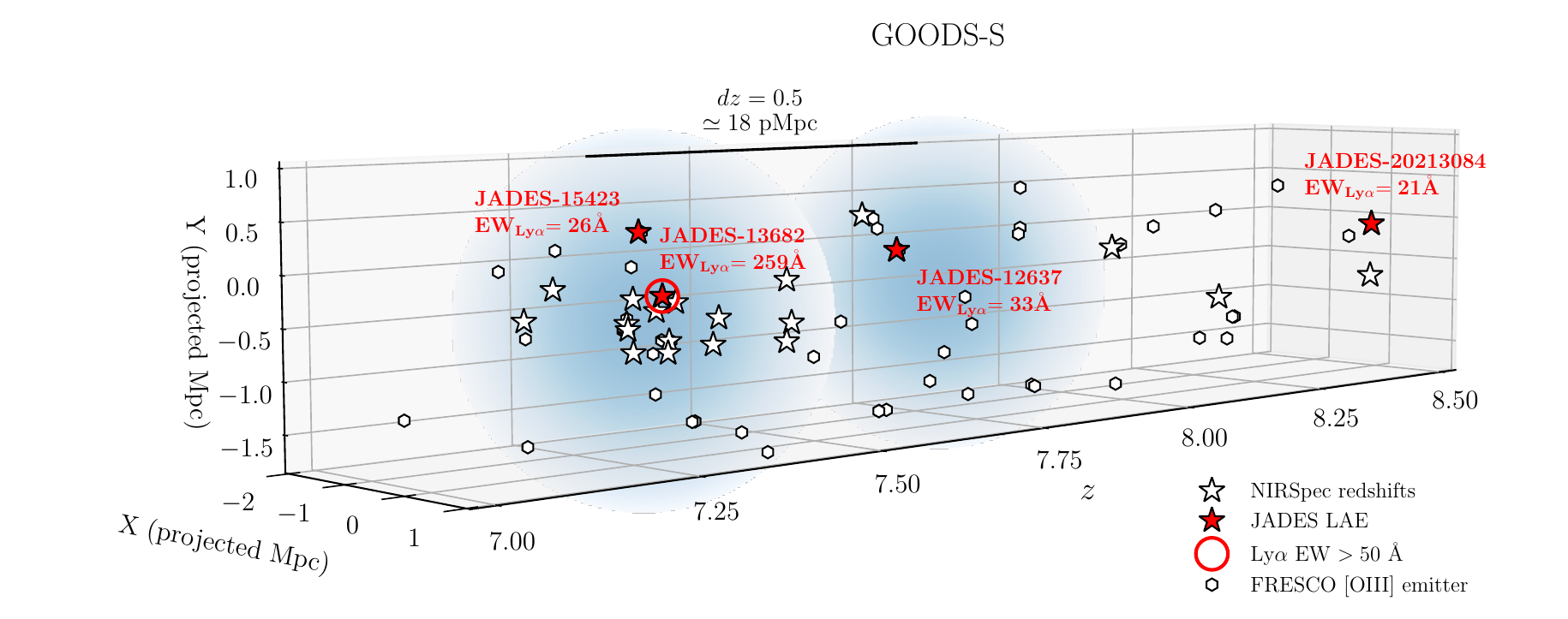}
\includegraphics[width=\linewidth]{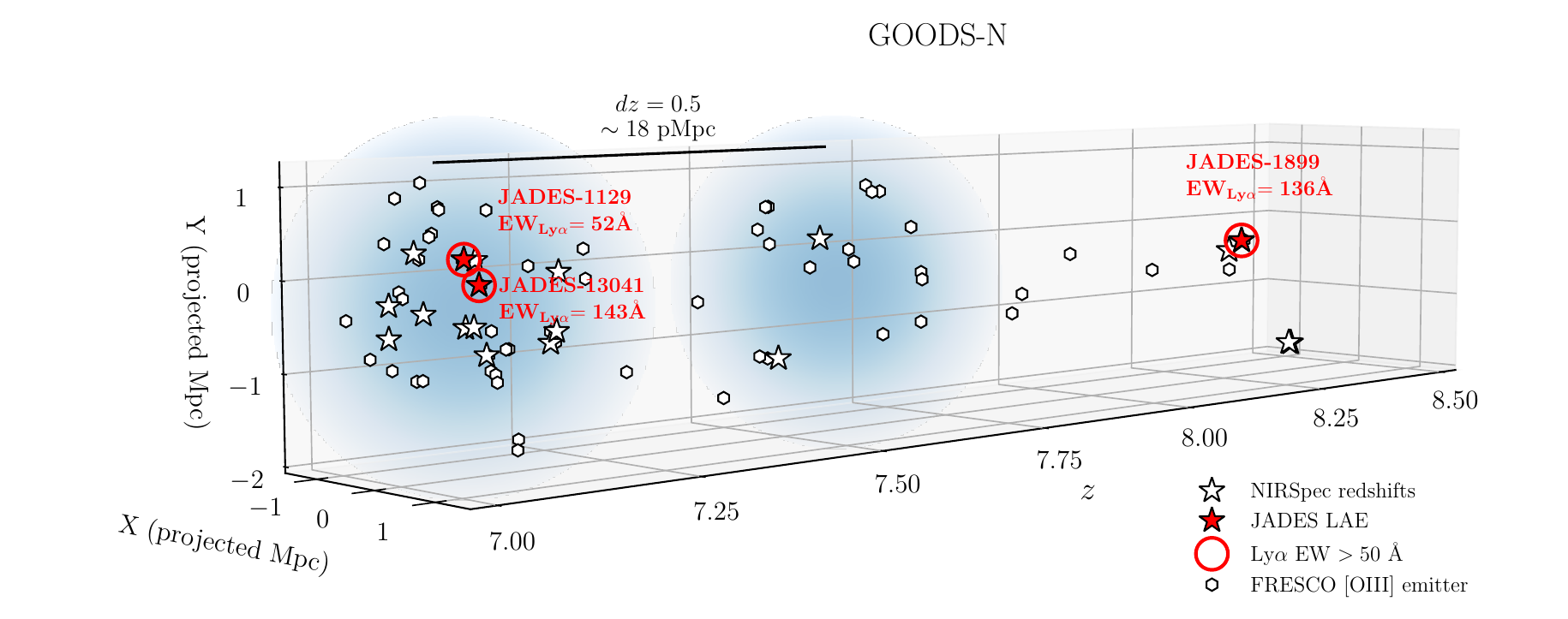}
\caption{Spatial distribution of spectroscopically confirmed galaxies in the two GOODS fields (top: GOODS-South; bottom: GOODS-North) at $z=7.0-8.5$ identified from both the JADES and the FRESCO dataset. We show galaxies with NIRSpec redshift confirmation as stars, and the [O~{\scriptsize III}] emitters identified from FRESCO observations \citep{Meyer2024} as hexagons. Ly$\alpha$ emitting galaxies are marked by red stars, and we highlight strong Ly$\alpha$ emitters with EW $>50$~\AA\ as red circles. We find possible overdensities of [O~{\scriptsize III}] emitters (blue shaded regions) around three strong Ly$\alpha$ emitters JADES-1129 ($z=7.087$) and JADES-13041 ($z=7.090$) in GOODS-N and JADES-13682 ($z=7.275$) in GOODS-S, and at $z\simeq7.7$ in GOODS-S and at $z\simeq7.6$ in GOODS-N.}
\label{fig:space_dist_jades}
\end{figure*}

We show the spatial distribution of spectroscopically-confirmed galaxies around the four strong Ly$\alpha$ emitters in Figure~\ref{fig:space_dist_jades}. 
To quantify overdensities, we first must measure the average number of [O~{\small III}] emitters in each GOODS field. 
We apply a flux limit of $F_{\rm [OIII]\lambda5007}>2\times10^{-18}$~erg~s$^{-1}$~cm$^{-2}$ (corresponding to S/N $>5$; \citealt{Oesch2023}) and measure the median number of [O~{\small III}] emitters in a large number ($N=1000$) redshift bins spanning $\Delta z=0.2$ (equivalent to line-of-sight distance of $\simeq6-8$~pMpc at $z\simeq7-8$) with randomly-chosen central redshifts between $z=7$ and $z=9$. 
For reference, the $62$~arcmin$^2$ area in each GOODS field has an effective radius of $R\simeq4.4$~arcmin (or $\simeq1.3$~pMpc at $z\sim7-9$), corresponding to a moderate-to-large bubble size at these redshifts \citep{Lu2024}. 
In each redshift bin, we find on average $2$ ($3$) [O~{\small III}]-detected galaxies in GOODS-S (GOODS-N). 
We will show below that these numbers are consistent with expectations given the UV luminosity function (LF) and luminosity-dependent [O~{\small III}]+H$\beta$ EW distribution at these redshifts. 

If the four strong $z\gtrsim7$ Ly$\alpha$ emitters in the FRESCO fields trace overdensities, we should see significantly more than $2-3$ galaxies in the $\Delta z=\pm0.1$ window centered on their redshifts. 
We find that there are $25$ and $8$ [O~{\small III}] emitters in the narrow redshift windows around JADES-1129 + JADES-13041 ($z\sim7.1$ in GOODS-N) and JADES-13682 ($z\sim7.3$ in GOODS-S), respectively (see Figure~\ref{fig:space_dist_jades}). 
These numbers indicate significant overdensities ($8\times$ for JADES-1129 + JADES-13041 and $4\times$ for JADES-13682) around both sources, as would be expected if these Ly$\alpha$ emitters trace regions with an excess of ionizing photons capable of carving out a sizable bubble. 
On the other hand, we find that the redshift bin around the strong Ly$\alpha$ emitter JADES-1899 contains no additional [O~{\small III}] emitters over the full GOODS-N field\footnote{There are several spectroscopically confirmed galaxies within $3$~pMpc of JADES-1899 (see Figure~\ref{fig:space_dist_jades}) which are not detected in FRESCO as they are either out of the footprint or fainter than the grism flux limit, but they are not abundant enough to suggest an overdensity.}. 
This suggests that this extremely strong Ly$\alpha$ emitter (EW $=136$~\AA) is able to transmit a large fraction of its line luminosity despite being situated in a potentially underdense region. 
Below we consider each of these environments in more detail.

JADES-13682 (first reported in \citealt{Saxena2023}) is one of the strongest Ly$\alpha$ emitters identified by \textit{JWST} to-date (EW $=259$~\AA), and a variety of studies have now confirmed that it is associated with an overdense region at $z\simeq7.3$ in GOODS-S \citep[e.g.,][]{Helton2023,Endsley2024,Witstok2024a}.
Our measurement of the spectroscopic overdensity ($4\times$) is broadly consistent with the findings in \citet{Witstok2024a}. 
They report an overdensity factor of $6.6\pm1.3$ based on analysis in \citet{Helton2023}. 
While slightly larger than our quoted overdensity, we note that they search over a considerably smaller area ($500$~pkpc $\times500$~pkpc) and a much smaller redshift range ($\Delta z\simeq0.014$), while also using a different grism redshift catalog (Sun et al., in prep.). 
A photometric overdensity ($\sim25\times$) has also been reported around JADES-13682 in \citet{Endsley2024}, leveraging medium bands sensitive to [O~{\small III}]+H$\beta$ emission. 
Our results demonstrate that the spectroscopic overdensity around this intense Ly$\alpha$ emitter spans a large physical volume (with effective radius $R\simeq1.3$~pMpc) in GOODS-S at $z\simeq7$. 
If this region is mostly ionized, we should expect to see the distribution of Ly$\alpha$ strengths enhanced (on average) throughout the overdensity. 
As is clear in Figure~\ref{fig:space_dist_jades}, one additional $z\simeq7.2$ galaxy in this overdensity is found with Ly$\alpha$ with EW $=26$~\AA\ (JADES-15423). 
Several other $z\simeq7.3$ galaxies in this area show non-detections of Ly$\alpha$, but the EW limits are not yet stringent for many of them.
Deeper spectroscopy will be required to robustly quantify the Ly$\alpha$ EW distribution in this volume. 

The spectroscopic overdensity in GOODS-N at $z\simeq7.1-7.2$ is one of the strongest known on large ($\gtrsim1$~pMpc) scales at $z\simeq7-8$. 
Two closely separated ($\simeq350$~pkpc) strong Ly$\alpha$ emitters, JADES-1129 (EW $=52$~\AA) and JADES-13041 (EW $=143$~\AA), are situated in this overdensity. 
They are actually the second and the third known Ly$\alpha$ emitters at this redshift in GOODS-N.
The first was reported over decade ago with a ground-based Ly$\alpha$ EW $=33$~\AA\ at $z=7.213$ \citep{Ono2012} and a separation of $5.9$~pMpc and $5.7$~pMpc from JADES-1129 and JADES-13041, respectively. 
The large-scale structure also hosts the massive dust-obscured AGN GNz7Q at $z=7.19$ \citep{Fujimoto2022,Meyer2024}. 
While some Ly$\alpha$ non-detections exist in the overdensity, statistics are not yet adequate to assess whether transmission is enhanced throughout the structure. 
If the Ly$\alpha$ transmission through the neutral IGM is equal to the Ly$\alpha$ escape fractions of the strong Ly$\alpha$ emitters ($\simeq0.5-0.6$), the typical distances between galaxies and the neutral IGM will be at least $1$~pMpc using the methods described in Section~\ref{sec:neutral_fraction}. 
Further spectroscopy will be able to estimate how the average Ly$\alpha$ transmission in this overdensity compares to the average transmission at $z\simeq 7$. 
This approach is possible with a modest investment in NIRSpec observations in this field (with an exposure time $\sim7$~hours) and should yield more robust measurements of the typical distances between galaxies and the neutral IGM. 

The absence of a strong overdensity around JADES-1899 in GOODS-N is puzzling, as its Ly$\alpha$ EW ($136$~\AA) is among the strongest known at these redshifts. 
Furthermore, its velocity profile reveals line emission near the systemic redshift (see \citealt{Tang2024b,Witstok2024b}), which is extremely rare at $z\gtrsim5$ \citep[e.g.,][]{Tang2024a} and likely indicates reduced IGM attenuation of the line profile. 
\citet{Witstok2024b} reported several galaxies that may be associated with the Ly$\alpha$ emitter (both photometric and spectroscopic) but concluded that the overdensity was likely not strong enough to power a large ($R\simeq3$~pMpc) ionized region around the Ly$\alpha$ emitter. 
Our results are consistent with this conclusion. 

There are two possible explanations for presence of strong Ly$\alpha$ without a significant population of surrounding [O~{\small III}] emitters.
First, it is conceivable that there is an overdensity present, but that it is not recovered by the [O~{\small III}] selections in FRESCO. 
This may be possible if the galaxies are weak in [O~{\small III}] emission.
We will consider this in more detail below. 
The second possibility is related to the hard radiation field of JADES-1899, as revealed by its strong emission from high-ionization species (N~{\small IV}], C~{\small IV}; \citealt{Tang2024b,Witstok2024b}). 
If the radiation field is intense enough to boost the local ionization fraction of hydrogen in its surroundings, it may reduce the impact of infalling IGM on resonant scattering of Ly$\alpha$ emission near line center\footnote{While the infalling IGM is ionized, the residual neutral hydrogen fraction is large enough at $z\gtrsim5$ to attenuate Ly$\alpha$ emission extending to the red side of the systemic redshift \citep[e.g.,][]{Santos2004,Dijkstra2007,Laursen2011,Mason2018}.} (see \citealt{Mason2020}). 
If the residual neutral hydrogen fraction in the infalling IGM is lower in the vicinity of JADES-1899, it would boost the transmission of the line at small velocities, helping to explain the velocity profile and the large EW. 
The damping wing from the neutral IGM outside the bubble would still attenuate the line emission, but the observed line profile could be explained with a smaller line-of-sight distance to the neutral IGM ($R\simeq1$ pMpc; \citealt{Tang2024b}), requiring a weaker overdensity that may be consistent with the FRESCO observations. 

The discussion above motivates the potential of combining \textit{JWST} measurements of galaxy environment with Ly$\alpha$ statistical distributions. 
Here we investigate the extent to which observations targeting [O~{\small III}] or H$\beta$ emission are able to recover overdensities, motivated by the absence of a strong overdensity in the field surrounding JADES-1899. 
We consider a volume at $z\simeq8$ with a radius of $1.3$~pMpc ($4.4$~arcmin) and calculate the average galaxy counts (as a function of M$_{\rm{UV}}$) expected in the volume according to the UV luminosity function (using the Schechter parameters in \citealt{Bouwens2021}). 
We then use measurements of the [O~{\small III}]+H$\beta$ EW distribution to assign each galaxy an [O~{\small III}] and H$\beta$ EW. 
We take into account measurements that indicate that the [O~{\small III}] and H$\beta$ EWs become weaker among fainter galaxies \citep{Endsley2024}, but we also consider cases where the rest-frame optical EW distributions are fixed with M$_{\rm{UV}}$. 
To convert from [O~{\small III}]+H$\beta$ to [O~{\small III}] on its own, we use a typical [O~{\small III}]~$\lambda5007$/H$\beta$ flux ratio ($\simeq6$) for reionization-era galaxies \citep[e.g.,][]{Nakajima2023,Sanders2023,Shapley2023,Tang2023}. 

\begin{figure}
\includegraphics[width=\linewidth]{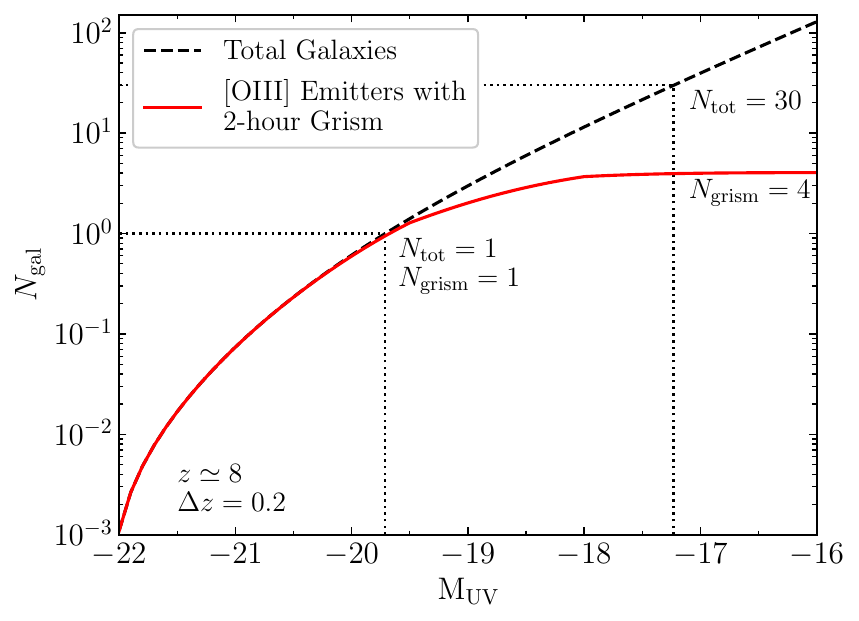}
\caption{Number of expected galaxies in a volume with an effective radius of $4.4$~arcmin (or $R=1.3$~pMpc, equivalent to the area of one of the FRESCO fields) and $\Delta z=0.2$. The black dashed line shows the total number of UV-selected $z\simeq8$ galaxies visible to a given M$_{\rm{UV}}$-limit, assuming an average density field according to the \citet{Bouwens2021} UV LFs. The red line shows the corresponding number of [O~{\small III}] emitters in the same area that would be recovered in $\simeq2$-hour grism observations down to the $5\sigma$ line flux limit ($>2\times10^{-18}$~erg~s$^{-1}$~cm$^{-2}$). Dotted lines highlight the absolute magnitude where we expect to identify $1$ and $30$ total galaxies in the volume. Existing grism observations become incomplete among the faint (M$_{\rm{UV}}>-19$) galaxy population. Fully characterizing overdensities will ultimately require deeper spectroscopy.}
\label{fig:Ngal_pred}
\end{figure}

We find that our chosen volume (with radius $R=1.3$~pMpc) should host $\simeq1$ ($\simeq30$) $z\simeq8$ galaxies to M$_{\rm{UV}}\simeq-19.5$ ($-17$; Figure~\ref{fig:Ngal_pred}). 
Of course the lower luminosity galaxies are most abundant by number, but their [O~{\small III}] and H$\beta$ lines are weaker (in line flux), making them more challenging to select. 
If we apply the same line flux limit we adopted for analysis of the FRESCO observations ($F_{\rm [OIII]\lambda5007}>2\times10^{-18}$~erg~s$^{-1}$~cm$^{-2}$), we find that we would primarily recover the small number of galaxies that are UV luminous in the continuum (see Figure~\ref{fig:Ngal_pred}). 
In particular, if we adopt the M$_{\rm{UV}}$-dependent [O~{\small III}] EW distributions \citep{Endsley2024}, we would only identify $4$ galaxies above the FRESCO flux limit, with typical continuum magnitudes in the range M$_{\rm{UV}}=-21$ to $-18$ (Figure~\ref{fig:Ngal_pred}). 
If we instead adopt a fixed [O~{\small III}] EW distribution, we find the recovered sample size is the same. 

Not surprisingly, moderate depth grism observations will be incomplete to the majority of fainter galaxies in the environment. 
This could be easily rectified with deeper NIRCam grism (or NIRSpec) observations of pointings found to be overdense. 
However, even at current $2$-hour depth, it is unlikely that a strong overdensity would be missed. 
For example, if the region within $R=1.3$~pMpc of JADES-1899 was $3\times$ overdense, we should have detected $12\pm3$ galaxies with $F_{\rm [OIII]\lambda5007}>2\times10^{-18}$~erg~s$^{-1}$~cm$^{-2}$, well above the $1$ that was detected. 
To explain the observations, we would require that the majority of the galaxies surrounding JADES-1899 happen to either be UV-faint or [O~{\small III}]-weak, both of which are not expected given our understanding of galaxy demographics. 
We thus conclude that the lack of strong overdensity around JADES-1899 is likely robust. 
This suggests that other factors (i.e., hard radiation field) may be permitting visibility of strong Ly$\alpha$. 
This underscores why Ly$\alpha$ statistical distributions (and not individual galaxy measurements) are required to infer the presence of large ionized bubbles in sub-regions of targeted fields. 


\section{Summary} \label{sec:summary}

\textit{JWST} has led to rapid progress in our characterization of Ly$\alpha$ emission from galaxies in the reionization era. 
The datasets are allowing improved constraints on the statistical distributions of Ly$\alpha$ emission, with the potential to greatly improve our knowledge of the early stages of reionization.
Here we present and analyze \textit{JWST}/NIRSpec Ly$\alpha$ spectra of $210$ galaxies at $6.5\lesssim z\lesssim13$.
We summarize our main results below. 

1. Using our reductions of publicly-available NIRSpec observations, we construct a sample of $210$ spectroscopically-confirmed galaxies at $z>6.5$ with Ly$\alpha$ constraints. 
These data come from observations targeting four independent fields: Abell 2744, EGS, GOODS-North and South. 
We measure the Ly$\alpha$ EWs and Ly$\alpha$ escape fractions for these $210$ galaxies, forming a sample with size that is $\sim3-6\times$ larger than early \textit{JWST} investigations of Ly$\alpha$ at $z>6.5$ \citep[e.g.,][]{Jones2024,Nakane2024,Napolitano2024}. 
Our sample contains $50$ galaxies at $z>8$, providing the largest $z>8$ sample to-date. 

2. The spectroscopic dataset includes $33$ $z>6.5$ galaxies with Ly$\alpha$ detections, with significant overlap with earlier papers in these fields \citep{Bunker2023a,Fujimoto2023b,Tang2023,Tang2024a,Tang2024b,Chen2024,Jones2024,Napolitano2024,Saxena2024,Witstok2024b}. 
Three of these source have yet to be identified in the literature (JADES-13041, JADES-14373, JADES-15423). 
In particular, JADES-13041 shows extremely strong Ly$\alpha$ emission (EW $=143$~\AA) in GOODS-N at a redshift ($z=7.1$) known to host two other Ly$\alpha$ emitters, hinting at a potential ionized structure at this redshift in GOODS-N. 

3. With improved statistics relative to previous studies, we quantify the distributions of Ly$\alpha$ EWs and Ly$\alpha$ escape fractions at $z>6.5$. 
We find that the fraction of galaxies showing large Ly$\alpha$ EW ($>25$~\AA) or large Ly$\alpha$ escape fraction ($f_{{\rm esc,Ly}\alpha}>0.2$) decreases by $4\times$ between $z\simeq5$ and $z\simeq9$, suggesting a rapid reduction in the transmission of Ly$\alpha$ photons.

4. If large ($\gtrsim1$~pMpc) ionized bubbles are in place at $z\gtrsim6.5$, we should find significant field-to-field variations in Ly$\alpha$ detections across the four survey fields. 
With our current dataset, we find that the Ly$\alpha$ emitter ($>$25~\AA) fraction in the EGS at $z=6.5-8.0$ ($26^{+10}_{-10}\%$) is two times larger than that measured in the other three fields. 
This field shows an elevated Ly$\alpha$ fraction, with little evidence for downward evolution from $z\simeq5$. 
While spectroscopic completeness in these fields is still limited, evidence continues to suggest that the EGS is likely to harbor large ionized regions at $z=6.5-8.0$ than the other three fields targeted with deep spectroscopy.

5. We consider implications for the IGM under the assumption that the declining Ly$\alpha$ transmission is driven by the neutral IGM. 
The results are consistent with a picture where the IGM grows increasingly neutral from $z=6.5-8.0$ ($x_{\rm HI}=0.48^{+0.15}_{-0.22}$) to $z=8.0-10.0$ ($x_{\rm HI}=0.81^{+0.12}_{-0.24}$) and $z=10.0-13.3$ ($x_{\rm HI}=0.89^{+0.08}_{-0.21}$). 
Whereas \textit{JWST} has indicated a surprisingly large density of UV photons at $z\gtrsim10$, Ly$\alpha$ observations have yet to reveal evidence that these galaxies made a significant impact on the ionization state of the IGM. 
Future observations are required to test this picture of early reionization with greater statistics. 

6. In addition to identifying likely ionized bubbles, \textit{JWST} is optimized for mapping the distribution of galaxies within these ionized regions. 
We use [O~{\small III}] emitters identified from the FRESCO grism spectroscopic survey \citep{Oesch2023,Meyer2024} to investigate the environments linked to the four strongest Ly$\alpha$ emitters (EW $>50$~\AA) at $z\gtrsim7$ in the FRESCO footprint. 
We find that three of the four strong Ly$\alpha$ emitters do trace significant ($\gtrsim4\times$) overdensities (JADES-1129 + JADES-13041 at $z\simeq7.1$ in GOODS-North and JADES-13682 at $z\simeq7.3$ in GOODS-South. 
However we do not find a strong overdensity around JADES-1899 ($z=8.3$, Ly$\alpha$ EW $=136$~\AA), suggesting it is possible to achieve large Ly$\alpha$ transmission in regions that are not overdense in [O~{\small III}] emitters. 
We suggest the hard radiation field of JADES-1899 may boost the transmission, allowing a large Ly$\alpha$ escape fraction with a smaller galaxy overdensity.


\section*{Acknowledgment}

The authors acknowledge the anonymous referee for the insightful comments, which improved the manuscript. 
We also thank St\'{e}phane Charlot and Jacopo Chevallard for providing access to the \texttt{BEAGLE} tool used for SED fitting analysis. 
MT acknowledges funding from the \textit{JWST} Arizona/Steward Postdoc in Early galaxies and Reionization (JASPER) Scholar contract at the University of Arizona. 
DPS acknowledges support from the National Science Foundation through the grant AST-2109066. 
CAM acknowledges support by the VILLUM FONDEN under grant 37459 and the Carlsberg Foundation under grant CF22-1322. 
The Cosmic Dawn Center (DAWN) is funded by the Danish National Research Foundation under grant DNRF140. 
RSE acknowledges generous support from the Peter and Patricia Gruber Foundation.

This work is based on observations made with the NASA/ESA/CSA \textit{James Webb Space Telescope}. 
The data were obtained from the Mikulski Archive for Space Telescopes at the Space Telescope Science Institute, which is operated by the Association of Universities for Research in Astronomy, Inc., under NASA contract NAS 5-03127 for \textit{JWST}. 
These observations are associated with program GTO 1180, 1181, 1210, and GO 3215 (JADES; doi: \href{http://archive.stsci.edu/doi/resolve/resolve.html?doi=10.17909/8tdj-8n28}{10.17909/8tdj-8n28}), ERS 1324 (GLASS; doi: \href{http://archive.stsci.edu/doi/resolve/resolve.html?doi=10.17909/kw3c-n857}{10.17909/kw3c-n857}), ERS 1345 and DDT 2750 (CEERS; doi: \href{http://archive.stsci.edu/doi/resolve/resolve.html?doi=10.17909/z7p0-8481}{10.17909/z7p0-8481}), as well as GO 2561 (UNCOVER). 
The authors acknowledge the JADES, GLASS, CEERS, and UNCOVER teams led by Daniel Eisenstein \& Nora L\"uetzgendorf, Tommaso Treu, Steven L. Finkelstein, Pablo Arrabal Haro, and Ivo Labb\'e \& Rachel Bezanson for developing their observing programs. 
This research is also based in part on observations made with the NASA/ESA \textit{Hubble Space Telescope} obtained from the Space Telescope Science Institute, which is operated by the Association of Universities for Research in Astronomy, Inc., under NASA contract NAS 5–26555. 
Part of the data products presented herein were retrieved from the Dawn \textit{JWST} Archive (DJA). 
DJA is an initiative of the Cosmic Dawn Center, which is funded by the Danish National Research Foundation under grant DNRF140. 
This work is based in part upon High Performance Computing (HPC) resources supported by the University of Arizona TRIF, UITS, and Research, Innovation, and Impact (RII) and maintained by the UArizona Research Technologies department. 

%

\vspace{5mm}


\software{\texttt{NumPy} \citep{Harris2020}, \texttt{Matplotlib} \citep{Hunter2007}, \texttt{SciPy} \citep{Virtanen2020}, \texttt{Astropy} \citep{AstropyCollaboration2013,AstropyCollaboration2018,AstropyCollaboration2022}}, \texttt{BEAGLE} \citep{Chevallard2016}, \texttt{Cloudy} \citep{Ferland2013}, \texttt{emcee} \citep{Foreman-Mackey2013}



\appendix
\restartappendixnumbering

\section{Table of L\lowercase{y}$\alpha$ Properties of $\lowercase{z}>6.5$ Galaxies}

\startlongtable
\centerwidetable
\begin{deluxetable*}{cccccccccccc}
\tablecaption{Ly$\alpha$ properties of galaxies at $z>6.5$ identified from the public \textit{JWST}/NIRSpec datasets.}
\tabletypesize{\scriptsize}
\tablehead{
ID & PID & R.A. & Decl. & M$_{\rm UV}$ & $z_{\rm spec}$ & $\Delta v^{\rm grating}_{{\rm Ly}\alpha}$ & EW$^{\rm grating}_{{\rm Ly}\alpha}$ & EW$^{\rm prism}_{{\rm Ly}\alpha}$ & $f^{\rm grating}_{{\rm esc,Ly}\alpha}$ & $f^{\rm prism}_{{\rm esc,Ly}\alpha}$ & Ref. \\
 & & (deg) & (deg) & (mag) & & (km s$^{-1}$) & (\AA) & (\AA) & & & 
}
\startdata
CEERS-662 & 2750 & $214.877890$ & $+52.897683$ & $-18.72$ & $6.533$ & & & $<30$ & & $<0.21$ & \\
CEERS-663 & 2750 & $214.878974$ & $+52.896755$ & $-20.06$ & $6.534$ & & & $<14$ & & $<0.18$ & (22) \\
CEERS-80575 & 1345 & $214.735573$ & $+52.749379$ & $-18.06$ & $6.534$ & & & $<54$ & & $<0.23$ & \\
JADES-71983 & 1181 & $189.197202$ & $+62.167059$ & $-19.43$ & $6.542$ & & $<40$ & $<54$ & $<0.91$ & $<0.68$ & \\
JADES-42722 & 1181 & $189.210631$ & $+62.163295$ & $-19.14$ & $6.544$ & & $<86$ & & $<0.97$ & & \\
JADES-57330 & 1181 & $189.228846$ & $+62.203997$ & $-20.73$ & $6.544$ & & $<11$ & $<21$ & $<0.16$ & $<0.41$ & \\
JADES-78455 & 1181 & $189.231977$ & $+62.202327$ & $-19.97$ & $6.548$ & & $<52$ & $<91$ & $<0.16$ & $<0.24$ & \\
JADES-58561 & 1181 & $189.221840$ & $+62.207361$ & $-19.30$ & $6.548$ & & $<17$ & $<37$ & $<0.09$ & $<0.15$ & \\
JADES-59734 & 1181 & $189.280691$ & $+62.210830$ & $-18.10$ & $6.548$ & & $<122$ & $<290$ & $<0.51$ & $<0.99$ & \\
JADES-78891 & 1181 & $189.225822$ & $+62.204212$ & $-21.06$ & $6.548$ & & $<13$ & $<16$ & $<0.04$ & $<0.06$ & \\
UNCOVER-38248 & 2561 & $3.610137$ & $-30.357605$ & $-17.89$ & $6.557$ & & & $<59$ & & $<0.20$ & \\
UNCOVER-37504 & 2561 & $3.610602$ & $-30.358963$ & $-17.73$ & $6.565$ & & & $<135$ & & $<0.30$ & \\
CEERS-1160 & 1345 & $214.805062$ & $+52.845888$ & $-19.81$ & $6.567$ & & & & & & (18) \\
CEERS-496 & 1345 & $214.864726$ & $+52.871733$ & $-18.98$ & $6.569$ & & & & & & \\
CEERS-265 & 1345 & $214.862818$ & $+52.889386$ & $-19.67$ & $6.570$ & & & $<28$ & & $<0.21$ & \\
JADES-17158 & 1180 & $53.150051$ & $-27.745009$ & $-20.38$ & $6.572$ & & $<37$ & $<56$ & $<0.12$ & $<0.25$ & \\
UNCOVER-6829 & 2561 & $3.593793$ & $-30.415422$ & $-19.04$ & $6.587$ & & & $<41$ & & $<0.41$ & \\
CEERS-781 & 2750 & $214.896623$ & $+52.895806$ & $-18.85$ & $6.590$ & & & & & & \\
JADES-1223 & 1181 & $189.120334$ & $+62.294871$ & $-19.56$ & $6.591$ & & $<17$ & & $<0.08$ & & \\
CEERS-386 & 1345 & $214.832162$ & $+52.885086$ & $-18.27$ & $6.614$ & & & $<186$ & & $<0.30$ & (18) \\
JADES-5447 & 1210 & $53.162884$ & $-27.769294$ & $-17.69$ & $6.622$ & & & $<44$ & & $<0.32$ & (10) \\
JADES-13607 & 1180 & $53.137425$ & $-27.765214$ & $-19.86$ & $6.623$ & & & $<45$ & & $<0.54$ & (26,32,35) \\
JADES-13647 & 1180 & $53.169513$ & $-27.753320$ & $-19.55$ & $6.625$ & & & $<45$ & & $<0.27$ & (26) \\
JADES-128261 & 3215 & $53.124281$ & $-27.777261$ & $-17.87$ & $6.626$ & & $<119$ & $<45$ & $<1.24$ & $<0.60$ & \\
JADES-58930 & 1180 & $53.105382$ & $-27.723466$ & $-21.71$ & $6.629$ & & $<8$ & & $<0.04$ & & (26) \\
JADES-9903 & 1210 & $53.169044$ & $-27.778836$ & $-18.63$ & $6.631$ & $377\pm101$ & $44\pm10$ & $56\pm4$ & $0.18\pm0.04$ & $0.21\pm0.02$ & (10,11,26,32,35) \\
JADES-8239 & 1181 & $189.248918$ & $+62.249738$ & $-18.67$ & $6.655$ & & $<25$ & $<188$ & $<0.50$ & $<0.35$ & \\
JADES-14373 & 1181 & $189.145785$ & $+62.273318$ & $-18.92$ & $6.663$ & $305\pm56$ & $49\pm15$ & $49\pm17$ & $0.38\pm0.12$ & $0.68\pm0.24$ & \\
JADES-18533 & 1181 & $189.121216$ & $+62.286409$ & $-19.47$ & $6.668$ & & & $<38$ & & $<0.34$ & \\
CEERS-1414 & 1345 & $215.128034$ & $+52.984949$ & $-20.87$ & $6.676$ & & & $32\pm4$ & & $0.07\pm0.01$ & (22,29) \\
JADES-44124 & 1181 & $189.272699$ & $+62.167408$ & $-19.65$ & $6.686$ & & $<26$ & $<57$ & $<0.10$ & $<0.17$ & \\
UNCOVER-8936 & 2561 & $3.604122$ & $-30.409544$ & $-17.02$ & $6.69^{\rm a}$ & & & $<81$ & & $<0.69$ & \\
CEERS-577 & 1345 & $214.892865$ & $+52.865164$ & $-18.52$ & $6.695$ & & & $<36$ & & $<0.15$ & (18) \\
JADES-39799 & 1181 & $189.263535$ & $+62.154791$ & $-19.43$ & $6.696$ & & $<27$ & $<48$ & $<0.57$ & $<0.31$ & \\
JADES-13286 & 1180 & $53.154958$ & $-27.815795$ & $-21.10$ & $6.707$ & & $<107$ & $<64$ & $<0.27$ & $<0.18$ & \\
JADES-38428 & 1181 & $189.179276$ & $+62.275902$ & $-21.49$ & $6.711$ & & $<8$ & $<20$ & $<0.06$ & $<0.12$ & \\
JADES-2113 & 1181 & $189.170327$ & $+62.229498$ & $-19.19$ & $6.713$ & & & $<62$ & & $<0.49$ & \\
UNCOVER-23619 & 2561 & $3.607272$ & $-30.380578$ & $-16.67$ & $6.717$ & & & $<178$ & & $<0.20$ & (30) \\
JADES-12359 & 1181 & $189.177807$ & $+62.265589$ & $-19.52$ & $6.718$ & & & $<68$ & & $<0.76$ & \\
JADES-89464 & 1181 & $189.186579$ & $+62.270901$ & $-21.00$ & $6.725$ & & $<11$ & $<24$ & $<0.11$ & $<0.17$ & \\
UNCOVER-43709 & 2561 & $3.575938$ & $-30.348029$ & $-18.35$ & $6.726$ & & & $<29$ & & $<0.26$ & \\
JADES-1948 & 1181 & $189.177306$ & $+62.291073$ & $-20.14$ & $6.728$ & & $<21$ & $<13$ & $<0.13$ & $<0.16$ & \\
CEERS-613 & 1345 & $214.882077$ & $+52.844347$ & $-19.16$ & $6.729$ & & & & & & \\
JADES-38420 & 1181 & $189.175117$ & $+62.282255$ & $-20.59$ & $6.733$ & $178\pm55$ & $51\pm6$ & $46\pm3$ & $0.34\pm0.04$ & $0.45\pm0.04$ & (34) \\
CEERS-81049 & 1345 & $214.789824$ & $+52.730790$ & $-19.74$ & $6.738$ & & & $99\pm3$ & & $0.38\pm0.01$ & (18,22,29) \\
JADES-28342 & 1181 & $189.224347$ & $+62.275586$ & $-20.37$ & $6.740$ & & & $40\pm13$ & & $0.18\pm0.06$ & \\
UNCOVER-2008 & 2561 & $3.592421$ & $-30.432827$ & $-17.34$ & $6.743$ & & & $<129$ & & $<0.22$ & \\
JADES-1076 & 1181 & $189.138504$ & $+62.275607$ & $-20.18$ & $6.743$ & & $<15$ & $<46$ & $<0.20$ & $<0.35$ & \\
CEERS-80925 & 1345 & $214.948699$ & $+52.853267$ & $-19.28$ & $6.746$ & & & $262\pm23$ & & $0.58\pm0.05$ & (22,29) \\
JADES-1066 & 1181 & $189.138071$ & $+62.274437$ & $-19.33$ & $6.747$ & & $<26$ & $<42$ & $<0.07$ & $<0.14$ & \\
UNCOVER-27335 & 2561 & $3.625075$ & $-30.375261$ & $-17.46$ & $6.755$ & & & $<285$ & & $<0.79$ & \\
JADES-926 & 1181 & $189.079811$ & $+62.256440$ & $-21.17$ & $6.756$ & & $<21$ & $<23$ & $<0.05$ & $<0.07$ & \\
JADES-896 & 1181 & $189.082664$ & $+62.252474$ & $-21.18$ & $6.759$ & & $<11$ & $<14$ & $<0.03$ & $<0.04$ & \\
JADES-10806 & 1181 & $189.153999$ & $+62.259541$ & $-18.52$ & $6.761$ & & $<22$ & $<159$ & $<0.81$ & $<1.25$ & \\
UNCOVER-36857 & 2561 & $3.582831$ & $-30.360296$ & $-18.41$ & $6.765$ & & & $<17$ & & $<0.20$ & \\
JADES-73977 & 1181 & $189.185500$ & $+62.179807$ & $-20.04$ & $6.766$ & & $<17$ & $<35$ & $<0.28$ & $<0.32$ & \\
UNCOVER-33673 & 2561 & $3.600114$ & $-30.365815$ & $-17.44$ & $6.772$ & & & $77\pm17$ & & $0.38\pm0.09$ & \\
JADES-13178 & 1180 & $53.118188$ & $-27.793013$ & $-19.32$ & $6.785$ & & $<36$ & $<65$ & $<0.20$ & $<0.43$ & (26) \\
CEERS-1064 & 1345 & $215.177170$ & $+53.048984$ & $-20.05$ & $6.790$ & & & $<68$ & & $<0.08$ & (18) \\
JADES-15362 & 1180 & $53.116314$ & $-27.761960$ & $-19.89$ & $6.793$ & & $<170$ & $<67$ & $<0.97$ & $<0.78$ & (26,32,35) \\
JADES-38681 & 1181 & $189.135639$ & $+62.263883$ & $-18.89$ & $6.800$ & & $<64$ & $<86$ & $<0.31$ & $<0.49$ & \\
JADES-18536 & 1181 & $189.155309$ & $+62.286471$ & $-20.20$ & $6.809$ & & & $<21$ & & $<0.11$ & \\
JADES-9104 & 1181 & $189.245274$ & $+62.252528$ & $-19.34$ & $6.816$ & & $<21$ & $<47$ & $<0.25$ & $<0.47$ & \\
JADES-17509 & 1180 & $53.147708$ & $-27.715370$ & $-20.54$ & $6.847$ & & & $<22$ & & $<0.29$ & (30) \\
JADES-137573 & 3215 & $53.156469$ & $-27.767261$ & $-17.59$ & $6.868$ & & $<80$ & $<52$ & $<1.44$ & $<0.20$ & \\
UNCOVER-11254 & 2561 & $3.580447$ & $-30.405021$ & $-18.69$ & $6.871$ & & & $15\pm2$ & & $0.05\pm0.01$ & (30) \\
UNCOVER-16155 & 2561 & $3.582957$ & $-30.395231$ & $-17.01$ & $6.874$ & & & $<51$ & & $<0.23$ & \\
UNCOVER-12899 & 2561 & $3.582353$ & $-30.402732$ & $-16.15$ & $6.877$ & & & $<115$ & & $<0.53$ & \\
JADES-8532 & 1180 & $53.145557$ & $-27.783795$ & $-19.86$ & $6.878$ & $335\pm54$ & $16\pm4$ & $<30$ & $0.18\pm0.05$ & $<0.29$ & (34) \\
JADES-1075 & 1181 & $189.202602$ & $+62.275513$ & $-19.84$ & $6.908$ & $216\pm54$ & $34\pm6$ & $43\pm9$ & $0.19\pm0.04$ & $0.46\pm0.10$ & (34) \\
CEERS-1143 & 1345 & $215.077023$ & $+52.969503$ & $-19.69$ & $6.928$ & & & $<10$ & & $<0.03$ & (18,20,22) \\
JADES-13609 & 1210 & $53.117297$ & $-27.764091$ & $-19.63$ & $6.929$ & & $<104$ & $<12$ & $<0.55$ & $<0.05$ & (10,11) \\
CEERS-481 & 1345 & $214.827786$ & $+52.850617$ & $-18.96$ & $6.930$ & & & $<34$ & & $<0.12$ & \\
CEERS-717 & 1345 & $215.081437$ & $+52.972181$ & $-21.42$ & $6.932$ & & & $<7$ & & $<0.05$ & (18,20,22) \\
CEERS-1142 & 1345 & $215.060750$ & $+52.958731$ & $-19.89$ & $6.938$ & & & $<20$ & & $<0.16$ & (18,29) \\
UNCOVER-9883 & 2561 & $3.549093$ & $-30.407617$ & $-17.18$ & $6.939$ & & & $<64$ & & $<0.84$ & \\
CEERS-716 & 1345 & $215.080361$ & $+52.993243$ & $-22.04$ & $6.962$ & & & $<8$ & & $<0.05$ & (18) \\
CEERS-445 & 2750 & $214.941607$ & $+52.929137$ & $-19.32$ & $6.980$ & & & $<39$ & & $<0.15$ & (22) \\
JADES-1254 & 1181 & $189.186285$ & $+62.225396$ & $-18.62$ & $6.990$ & & $<44$ & $<199$ & $<0.62$ & $<0.65$ & \\
CEERS-1102 & 1345 & $215.091036$ & $+52.954300$ & $-19.28$ & $6.992$ & & & $<13$ & & $<0.18$ & (18,20) \\
JADES-1893 & 1181 & $189.205296$ & $+62.250776$ & $-20.40$ & $6.994$ & & $<30$ & $<39$ & $<0.25$ & $<0.34$ & \\
JADES-7424 & 1181 & $189.232904$ & $+62.247376$ & $-19.01$ & $6.996$ & & $<27$ & $<59$ & $<0.18$ & $<0.43$ & \\
JADES-2316 & 1181 & $189.162535$ & $+62.258238$ & $-19.36$ & $6.998$ & & $<19$ & $<24$ & $<0.40$ & $<0.17$ & \\
CEERS-80244 & 1345 & $214.902160$ & $+52.869762$ & $-18.42$ & $7.001$ & & & $<31$ & & $<0.25$ & \\
JADES-5088 & 1181 & $189.172523$ & $+62.240540$ & $-18.42$ & $7.001$ & & & $<160$ & & $<0.48$ & \\
JADES-1166 & 1181 & $189.183348$ & $+62.287738$ & $-19.99$ & $7.025$ & & $<12$ & $<34$ & $<0.22$ & $<0.32$ & \\
CEERS-407 & 1345 & $214.839298$ & $+52.882573$ & $-18.78$ & $7.029$ & & & & & & (18,20) \\
CEERS-80401 & 1345 & $214.944401$ & $+52.837599$ & $-17.60$ & $7.032$ & & & $<28$ & & $<0.12$ & \\
UNCOVER-28876 & 2561 & $3.569594$ & $-30.373222$ & $-17.27$ & $7.034$ & & & $<88$ & & $<0.23$ & (30) \\
CEERS-542 & 1345 & $214.831620$ & $+52.831500$ & $-19.94$ & $7.061$ & & & $<23$ & & $<0.33$ & (18,28) \\
JADES-40307 & 1181 & $189.042940$ & $+62.251496$ & $-19.47$ & $7.078$ & & $<56$ & $<109$ & $<0.41$ & $<0.78$ & \\
JADES-1129 & 1181 & $189.179796$ & $+62.282395$ & $-19.44$ & $7.087$ & $122\pm53$ & $52\pm9$ & $70\pm13$ & $0.20\pm0.04$ & $0.26\pm0.05$ & (34) \\
CEERS-749 & 1345 & $215.002841$ & $+53.007593$ & $-18.55$ & $7.088$ & & & $<44$ & & $<0.13$ & \\
JADES-7675 & 1181 & $189.096301$ & $+62.247978$ & $-17.46$ & $7.088$ & & & & & & \\
JADES-13041 & 1181 & $189.203769$ & $+62.268425$ & $-19.15$ & $7.090$ & $233\pm53$ & $143\pm7$ & $133\pm12$ & $0.57\pm0.03$ & $0.46\pm0.04$ & \\
JADES-1936 & 1181 & $189.195710$ & $+62.282424$ & $-19.50$ & $7.090$ & & & $<51$ & & $<0.35$ & \\
CEERS-44 & 1345 & $215.001118$ & $+53.011274$ & $-19.78$ & $7.104$ & & & $83\pm4$ & & $0.71\pm0.04$ & (16,18,20,28,29) \\
CEERS-534 & 1345 & $214.859117$ & $+52.853639$ & $-19.87$ & $7.115$ & & & $<21$ & & $<0.16$ & (18,28) \\
UNCOVER-60141 & 2561 & $3.620339$ & $-30.388599$ & $-17.04$ & $7.128$ & & & $<102$ & & $<0.34$ & \\
JADES-3982 & 1181 & $189.109421$ & $+62.238795$ & $-19.56$ & $7.132$ & & $<25$ & $<51$ & $<0.19$ & $<0.46$ & \\
JADES-9442 & 1180 & $53.138054$ & $-27.781868$ & $-18.11$ & $7.136$ & & $<278$ & $<157$ & $<1.24$ & $<0.71$ & (26) \\
JADES-66336 & 1181 & $189.259285$ & $+62.235461$ & $-19.29$ & $7.140$ & & $<56$ & $<43$ & $<0.10$ & $<0.20$ & \\
JADES-24819 & 1181 & $189.136488$ & $+62.223395$ & $-21.13$ & $7.140$ & & $<27$ & & $<0.11$ & & \\
JADES-4530 & 1181 & $189.109146$ & $+62.238656$ & $-18.90$ & $7.142$ & & & $<95$ & & $<0.77$ & \\
JADES-67006 & 1181 & $189.249820$ & $+62.241215$ & $-20.45$ & $7.154$ & & & $<27$ & & $<0.10$ & \\
CEERS-499 & 1345 & $214.813004$ & $+52.834170$ & $-17.70$ & $7.168$ & & $<70$ & & $<0.46$ & & (20) \\
CEERS-829 & 1345 & $214.861585$ & $+52.876166$ & $-19.37$ & $7.172$ & & & $<21$ & & $<0.14$ & (16,18,28) \\
CEERS-80374 & 1345 & $214.898085$ & $+52.824897$ & $-18.34$ & $7.174$ & & & $201\pm15$ & & $0.25\pm0.02$ & (18,22,28,29) \\
CEERS-498 & 1345 & $214.813048$ & $+52.834234$ & $-20.31$ & $7.178$ & & & $33\pm2$ & & $0.16\pm0.01$ & (18,20,28,29) \\
CEERS-439 & 1345 & $214.825351$ & $+52.863063$ & $-19.14$ & $7.180$ & & & $71\pm7$ & & $0.23\pm0.02$ & (16,18,28,29) \\
CEERS-1038 & 1345 & $215.039712$ & $+52.901596$ & $-19.37$ & $7.194$ & & $<161$ & & $<0.32$ & & (18,20) \\
JADES-13905 & 1210 & $53.118332$ & $-27.769014$ & $-18.91$ & $7.198$ & & $<22$ & $<22$ & $<0.43$ & $<0.13$ & (10,11) \\
JADES-11547 & 1180 & $53.164826$ & $-27.788260$ & $-20.12$ & $7.234$ & & $<43$ & $<96$ & $<0.26$ & $<0.42$ & (26) \\
JADES-9942 & 1180 & $53.161711$ & $-27.785395$ & $-19.99$ & $7.236$ & & $<91$ & $<258$ & $<0.44$ & $<0.18$ & (26) \\
JADES-9886 & 1180 & $53.165547$ & $-27.772675$ & $-18.42$ & $7.239$ & & $<36$ & $<110$ & $<0.24$ & $<0.63$ & (26) \\
JADES-27058 & 1181 & $189.124743$ & $+62.268568$ & $-19.24$ & $7.24^{\rm a}$ & & $<39$ & $<55$ & & & \\
JADES-15423 & 1180 & $53.169576$ & $-27.738063$ & $-20.07$ & $7.242$ & $269\pm52$ & $26\pm3$ & $<31$ & $0.24\pm0.03$ & $<0.24$ & \\
JADES-13729 & 1180 & $53.182035$ & $-27.778060$ & $-16.99$ & $7.248$ & & $<61$ & $<202$ & $<0.43$ & $<1.16$ & \\
JADES-5115 & 1210 & $53.152841$ & $-27.801944$ & $-17.92$ & $7.256$ & & $<102$ & $<47$ & $<1.36$ & $<0.63$ & (10) \\
JADES-13173 & 1180 & $53.183937$ & $-27.799990$ & $-19.26$ & $7.260$ & & $<32$ & $<104$ & $<0.28$ & $<0.46$ & \\
JADES-2958 & 1180 & $53.183750$ & $-27.793891$ & $-19.35$ & $7.262$ & & & $<84$ & & $<0.35$ & \\
JADES-13682 & 1210 & $53.167453$ & $-27.772035$ & $-17.60$ & $7.275$ & $217\pm93$ & $259\pm54$ & $371\pm28$ & $0.53\pm0.11$ & $0.63\pm0.05$ & (10,11,19,26,35) \\
JADES-9425 & 1180 & $53.179755$ & $-27.774648$ & $-19.20$ & $7.276$ & & $<68$ & $<205$ & $<0.17$ & $<0.51$ & \\
GLASS-10021 & 1324 & $3.608517$ & $-30.418542$ & $-21.15$ & $7.287$ & & $<12$ & & $<0.02$ & & (18) \\
UNCOVER-8669 & 2561 & $3.553777$ & $-30.410131$ & $-18.41$ & $7.296$ & & & $<353$ & & $<0.87$ & \\
JADES-9315 & 1210 & $53.155086$ & $-27.801774$ & $-19.16$ & $7.36^{\rm a}$ & & $<78$ & $<12$ & $<1.04$ & $<0.51$ & (10) \\
JADES-11541 & 1180 & $53.149414$ & $-27.788265$ & $-18.05$ & $7.376$ & & $<171$ & $<226$ & $<0.17$ & $<0.66$ & \\
JADES-13552 & 1180 & $53.183429$ & $-27.790971$ & $-19.99$ & $7.429$ & & & $<32$ & & $<0.21$ & \\
JADES-60331 & 1181 & $189.275238$ & $+62.212439$ & $-18.91$ & $7.431$ & & $<22$ & $<56$ & $<0.28$ & $<0.59$ & \\
CEERS-52 & 1345 & $215.011631$ & $+53.014149$ & $-18.95$ & $7.434$ & & & $<69$ & & $<0.17$ & \\
CEERS-1163 & 1345 & $214.990478$ & $+52.971998$ & $-20.78$ & $7.447$ & & $<123$ & $<5$ & $<0.50$ & $<0.05$ & (16,18,20) \\
CEERS-38 & 1345 & $214.994942$ & $+53.007923$ & $-19.83$ & $7.451$ & & $<23$ & & $<0.17$ & & \\
CEERS-698 & 1345 & $215.050341$ & $+53.007447$ & $-21.70$ & $7.470$ & $534\pm91$ & $9\pm2$ & & $0.04\pm0.01$ & & (3,4,18,20,27,31) \\
JADES-30083556 & 3215 & $53.147356$ & $-27.805434$ & $-17.78$ & $7.474$ & & $<141$ & $<63$ & $<1.50$ & $<0.70$ & \\
CEERS-80432 & 1345 & $214.812056$ & $+52.746745$ & $-20.11$ & $7.475$ & & & $53\pm3$ & & $0.21\pm0.01$ & (18,22,28,29) \\
CEERS-80372 & 1345 & $214.927820$ & $+52.850001$ & $-19.19$ & $7.483$ & & & $<14$ & & $<0.05$ & (18,29) \\
CEERS-434 & 2750 & $214.898022$ & $+52.892971$ & $-18.99$ & $7.485$ & & & & & & (30) \\
CEERS-80239 & 1345 & $214.896056$ & $+52.869858$ & $-18.21$ & $7.487$ & & & $219\pm17$ & & $0.62\pm0.05$ & (18,22) \\
CEERS-80445 & 1345 & $214.843083$ & $+52.747880$ & $-20.77$ & $7.508$ & & & $51\pm1$ & & $0.39\pm0.01$ & (18,28,29) \\
CEERS-689 & 1345 & $214.998853$ & $+52.942090$ & $-21.08$ & $7.545$ & & $<80$ & & $<0.08$ & & (18,20,27) \\
CEERS-449 & 2750 & $214.940484$ & $+52.932557$ & $-18.33$ & $7.551$ & & & $<40$ & & $<0.18$ & \\
JADES-44323 & 1180 & $53.167790$ & $-27.736167$ & $-20.22$ & $7.555$ & & $<60$ & $<87$ & $<0.48$ & $<0.34$ & \\
JADES-17038 & 1180 & $53.087229$ & $-27.777053$ & $-19.55$ & $7.560$ & & $<35$ & $<126$ & $<0.79$ & $<0.60$ & (26) \\
UNCOVER-38059 & 2561 & $3.605255$ & $-30.357940$ & $-18.89$ & $7.585$ & & & $<30$ & & $<0.21$ & \\
JADES-38684 & 1181 & $189.121094$ & $+62.277823$ & $-19.77$ & $7.61^{\rm a}$ & & $<61$ & $<144$ & & & \\
CEERS-80025 & 1345 & $214.806074$ & $+52.750864$ & $-19.68$ & $7.657$ & & & $<13$ & & $<0.06$ & (7) \\
JADES-12637 & 1180 & $53.133465$ & $-27.760387$ & $-20.72$ & $7.660$ & $277\pm49$ & $33\pm3$ & $<40$ & $0.22\pm0.03$ & $<0.23$ & (26,32,35) \\
UNCOVER-18924 & 2561 & $3.581046$ & $-30.389559$ & $-15.09$ & $7.686$ & & & $<102$ & & $<0.34$ & \\
CEERS-686 & 1345 & $215.150886$ & $+52.989555$ & $-20.02$ & $7.750$ & & & $44\pm3$ & & $0.22\pm0.02$ & (18,20,27,29) \\
CEERS-20 & 1345 & $214.830660$ & $+52.887775$ & $-18.94$ & $7.762$ & & $<47$ & $<195$ & $<1.07$ & $<0.13$ & (14) \\
CEERS-1023 & 1345 & $215.188414$ & $+53.033652$ & $-20.87$ & $7.776$ & & $<25$ & $<9$ & $<0.20$ & $<0.04$ & (16,18,20) \\
CEERS-1027 & 1345 & $214.882999$ & $+52.840418$ & $-20.73$ & $7.819$ & $312\pm88$ & $26\pm3$ & & $0.11\pm0.02$ & & (16,18,20,31) \\
GLASS-100001$^{\rm b}$ & 1324 & $3.603888$ & $-30.382263$ & $-20.53$ & $7.874$ & & $<11$ & $<26$ & $<0.18$ & $<0.25$ & (16,17,18,22) \\
GLASS-100003 & 1324 & $3.604514$ & $-30.380444$ & $-20.49$ & $7.877$ & & $<9$ & & $<0.06$ & & (17,18) \\
GLASS-100005 & 1324 & $3.606455$ & $-30.380977$ & $-20.01$ & $7.879$ & & $<9$ & & $<0.24$ & & (17,18) \\
GLASS-10000$^{\rm b}$ & 1324 & $3.601340$ & $-30.379204$ & $-20.09$ & $7.881$ & & $<2$ & $<15$ & $<0.05$ & $<0.06$ & (17,18,22) \\
UNCOVER-23604 & 2561 & $3.605247$ & $-30.380584$ & $-17.85$ & $7.883$ & & & $43\pm5$ & & $0.16\pm0.02$ & (22) \\
CEERS-355 & 2750 & $214.944758$ & $+52.931456$ & $-19.14$ & $7.925$ & & & $<27$ & & $<0.27$ & (8) \\
JADES-5173 & 1210 & $53.156827$ & $-27.767163$ & $-18.91$ & $7.981$ & & $<18$ & $<20$ & $<0.10$ & $<0.07$ & (10,11,26,32,35) \\
CEERS-4 & 1345 & $215.005365$ & $+52.996697$ & $-18.80$ & $7.992$ & & $<48$ & & $<0.40$ & & (14,18,20) \\
CEERS-3 & 1345 & $215.005183$ & $+52.996582$ & $-18.81$ & $8.006$ & & & $<86$ & & $<0.07$ & (14,18,20,28) \\
CEERS-1149 & 1345 & $215.089737$ & $+52.966189$ & $-20.42$ & $8.175$ & & $<54$ & $<10$ & $<0.11$ & $<0.03$ & (16,18,20,31) \\
JADES-20198852 & 3215 & $53.107765$ & $-27.812936$ & $-19.15$ & $8.268$ & & $<72$ & $<20$ & $<1.45$ & $<0.44$ & \\
JADES-1899 & 1181 & $189.197727$ & $+62.256965$ & $-19.42$ & $8.279$ & $32\pm46$ & $136\pm9$ & $108\pm11$ & $0.38\pm0.03$ & $0.41\pm0.04$ & (34,36) \\
JADES-45131 & 1181 & $189.211388$ & $+62.170301$ & $-19.68$ & $8.368$ & & $<37$ & $<43$ & $<0.11$ & $<0.29$ & \\
JADES-45170 & 1181 & $189.207167$ & $+62.170389$ & $-17.79$ & $8.368$ & & $<64$ & $<60$ & $<0.23$ & $<0.41$ & \\
JADES-5776 & 1181 & $189.077273$ & $+62.242534$ & $-19.20$ & $8.369$ & & & $<205$ & & $<0.41$ & \\
JADES-6139 & 1210 & $53.164469$ & $-27.802184$ & $-18.18$ & $8.473$ & & $<143$ & $<174$ & $<0.14$ & $<0.23$ & (10,11) \\
JADES-20213084 & 3215 & $53.158910$ & $-27.765077$ & $-19.09$ & $8.486$ & $154\pm81$ & $21\pm3$ & $26\pm3$ & $0.14\pm0.02$ & $0.20\pm0.03$ & (33,36) \\
UNCOVER-10646 & 2561 & $3.636963$ & $-30.406362$ & $-21.44$ & $8.511$ & & & $13\pm1$ & & $0.04\pm0.01$ & (15) \\
JADES-74111 & 1181 & $189.180490$ & $+62.180463$ & $-20.59$ & $8.605$ & & $<13$ & $<54$ & $<0.70$ & $<0.48$ & \\
CEERS-1029 & 1345 & $215.218788$ & $+53.069869$ & $-21.53$ & $8.610$ & $2100\pm80$ & $3\pm1$ & $<4$ & $0.08\pm0.02$ & $<0.09$ & (5,16,18,20,28) \\
CEERS-80083 & 1345 & $214.961281$ & $+52.842352$ & $-19.37$ & $8.638$ & & & $<31$ & & $<0.10$ & (7,18) \\
JADES-54165 & 1181 & $189.271837$ & $+62.195178$ & $-19.52$ & $8.657$ & & $<29$ & $<36$ & & & (30) \\
CEERS-1019 & 1345 & $215.035388$ & $+52.890671$ & $-22.09$ & $8.678$ & $288\pm80$ & $3\pm1$ & & $0.02\pm0.01$ & & (1,6,16,18,20,31) \\
CEERS-1025 & 1345 & $214.967526$ & $+52.932953$ & $-21.08$ & $8.714$ & & & & & & (18,20) \\
JADES-20100293 & 3215 & $53.168738$ & $-27.816978$ & $-17.60$ & $8.749$ & & $<197$ & $<110$ & $<0.85$ & $<0.62$ & \\
CEERS-28 & 2750 & $214.938633$ & $+52.911750$ & $-20.67$ & $8.763$ & & & $<15$ & & $<0.13$ & (8,25) \\
CEERS-2 & 1345 & $214.994398$ & $+52.989382$ & $-20.17$ & $8.809$ & & & & & & (14) \\
JADES-20111790 & 3215 & $53.116857$ & $-27.800565$ & $-18.10$ & $8.822$ & & $<64$ & $<142$ & $<0.77$ & $<0.51$ & \\
CEERS-7 & 1345 & $215.011709$ & $+52.988306$ & $-20.55$ & $8.866$ & & $<27$ & $<21$ & $<0.43$ & $<0.23$ & (14,18) \\
CEERS-23 & 1345 & $214.901252$ & $+52.847000$ & $-19.02$ & $8.880$ & & $<20$ & & $<0.23$ & & (14,20) \\
JADES-20110306 & 3215 & $53.169132$ & $-27.802920$ & $-17.91$ & $8.919$ & & & $<47$ & & $<1.06$ & \\
JADES-13643 & 1181 & $189.204169$ & $+62.270759$ & $-19.13$ & $8.930$ & & $<45$ & $<239$ & $<0.23$ & $<0.59$ & \\
CEERS-24 & 1345 & $214.897232$ & $+52.843858$ & $-19.64$ & $8.999$ & & $<21$ & & $<0.08$ & & (14,20) \\
JADES-619 & 1181 & $189.158251$ & $+62.221361$ & $-19.99$ & $9.070$ & & $<24$ & $<42$ & $<0.13$ & $<0.50$ & \\
JADES-17858 & 1181 & $189.142208$ & $+62.284594$ & $-19.89$ & $9.209$ & & $<11$ & $<76$ & $<0.12$ & $<0.81$ & \\
JADES-19715 & 1181 & $189.138322$ & $+62.289869$ & $-20.13$ & $9.305$ & & & $<49$ & & $<0.42$ & \\
UNCOVER-3686 & 2561 & $3.617199$ & $-30.425536$ & $-21.71$ & $9.321$ & & & $<5$ & & $<0.13$ & (15) \\
JADES-3990 & 1181 & $189.016992$ & $+62.241585$ & $-20.69$ & $9.380$ & & & $<29$ & & $<0.08$ & \\
JADES-265801 & 3215 & $53.112436$ & $-27.774619$ & $-20.29$ & $9.433$ & & & $<3$ & & $<0.03$ & (10,11,23) \\
UNCOVER-22223 & 2561 & $3.568115$ & $-30.383051$ & $-16.65$ & $9.566$ & & & $<15$ & & $<0.07$ & (15) \\
JADES-59720 & 1181 & $189.239795$ & $+62.210830$ & $-19.70$ & $9.633$ & & & $<79$ & & $<0.66$ & \\
JADES-80088 & 1181 & $189.239122$ & $+62.210934$ & $-20.15$ & $9.737$ & & & $<84$ & & $<0.39$ & \\
JADES-55757 & 1181 & $189.217683$ & $+62.199490$ & $-19.86$ & $9.742$ & & & $<31$ & & $<0.18$ & \\
CEERS-80026 & 1345 & $214.811848$ & $+52.737113$ & $-20.09$ & $9.75^{\rm a}$ & & & $<10$ & & & (7) \\
UNCOVER-13151 & 2561 & $3.592505$ & $-30.401463$ & $-17.46$ & $9.803$ & & & $<19$ & & $<0.24$ & (15) \\
JADES-11508 & 1181 & $189.184460$ & $+62.262491$ & $-19.64$ & $9.933$ & & & $<84$ & & $<0.21$ & \\
UNCOVER-26185 & 2561 & $3.567071$ & $-30.377862$ & $-18.93$ & $10.065$ & & & $<16$ & & $<0.10$ & (15) \\
CEERS-64 & 2750 & $214.922774$ & $+52.911525$ & $-19.50$ & $10.07^{\rm a}$ & & & $<30$ & & & (8) \\
CEERS-80041 & 1345 & $214.732534$ & $+52.758092$ & $-20.37$ & $10.23^{\rm a}$ & & & $<14$ & & & (7) \\
UNCOVER-37126 & 2561 & $3.590110$ & $-30.359742$ & $-19.80$ & $10.39^{\rm a}$ & & & $<11$ & & & (15) \\
JADES-14177 & 1210 & $53.158837$ & $-27.773500$ & $-18.39$ & $10.39^{\rm a}$ & & & $<35$ & & & (10,12) \\
JADES-3991 & 1181 & $189.106056$ & $+62.242052$ & $-21.89$ & $10.604$ & $633\pm37$ & $12\pm2$ & $<21$ & $0.03\pm0.01$ & $<0.09$ & (2,9) \\
CEERS-10 & 2750 & $214.906630$ & $+52.945507$ & $-20.18$ & $11.39^{\rm a}$ & & & $<38$ & & & (8) \\
JADES-14220 & 1210 & $53.164768$ & $-27.774627$ & $-19.45$ & $11.55^{\rm a}$ & & & $<45$ & & & (10,12,24) \\
CEERS-1 & 2750 & $214.943138$ & $+52.942444$ & $-20.02$ & $11.55^{\rm a}$ & & & $<26$ & & & (8,24) \\
UNCOVER-38766 & 2561 & $3.513562$ & $-30.356798$ & $-18.91$ & $12.39^{\rm a}$ & & & $<32$ & & & (15,21) \\
JADES-20096216 & 3215 & $53.166346$ & $-27.821558$ & $-18.98$ & $12.513$ & & & $<103$ & & & (10,12,13) \\
JADES-20128771 & 3215 & $53.149886$ & $-27.776504$ & $-18.70$ & $13.22^{\rm a}$ & & & $<45$ & & & (10,12,24) \\
\enddata
\tabletypesize{\small}
\tablecomments{We list Ly$\alpha$ EWs (EW$_{{\rm Ly}\alpha}$) and Ly$\alpha$ escape fractions ($f_{{\rm esc,Ly}\alpha}$) derived from both the NIRSpec grating and prism spectra. Ly$\alpha$ velocity offsets ($\Delta v_{{\rm Ly}\alpha}$) are derived from grating spectra. PID is the \textit{JWST} program ID. Absolute UV magnitudes of galaxies in GLASS or UNCOVER are corrected for magnifications using models in \citet{Furtak2023b}. Spectroscopic redshifts ($z_{\rm spec}$) listed are the systemic redshifts measured from grating spectra, or from prism spectra for galaxies without grating spectra. Ly$\alpha$ escape fractions are computed assuming case B recombination. a: Redshifts derived from Ly$\alpha$ break; b: GLASS-100001 and GLASS-10000 have been observed in both GLASS (with high resolution grating spectra) and UNCOVER (with low resolution prism spectra) programs, their NIRSpec IDs in the UNCOVER dataset are UNCOVER-60157 and UNCOVER-24531, respectively.\\
{\bf Reference}: (1) \citet{Zitrin2015}, (2) \citet{Oesch2016}, (3) \citet{Roberts-Borsani2016}, (4) \citet{Stark2017}, (5) \citet{Larson2022}, (6) \citet{Larson2023}, (7) \citet{ArrabalHaro2023a}, (8) \citet{ArrabalHaro2023b}, (9) \citet{Bunker2023a}, (10) \citet{Bunker2023b}, (11) \citet{Cameron2023}, (12) \citet{Curtis-Lake2023}, (13) \citet{DEugenio2023}, (14) \citet{Fujimoto2023a}, (15) \citet{Fujimoto2023b}, (16) \citet{Heintz2023}, (17) \citet{Morishita2023}, (18) \citet{Nakajima2023}, (19) \citet{Saxena2023}, (20) \citet{Tang2023}, (21) \citet{Wang2023}, (22) \citet{Chen2024}, (23) \citet{Curti2024}, (24) \citet{Hainline2024}, (25) \citet{Heintz2024b}, (26) \citet{Jones2024}, (27) \citet{Jung2024}, (28) \citet{Nakane2024}, (29) \citet{Napolitano2024}, (30) \citet{Roberts-Borsani2024}, (31) \citet{Sanders2024}, (32) \citet{Saxena2024}, (33) \citet{Tang2024a}, (34) \citet{Tang2024b}, (35) \citet{Witstok2024a}, (36) \citet{Witstok2024b}.}
\label{tab:sources}
\end{deluxetable*}


\bibliography{NIRSpec_zg6_Lya}{}
\bibliographystyle{aasjournal}



\end{document}